\documentclass[%
superscriptaddress,
preprint,
preprintnumbers,
 amsmath,amssymb,
 aps,
 pra,
floatfix,
]{revtex4-1}

\usepackage{graphicx}% Include figure files
\usepackage{dcolumn}% Align table columns on decimal point
\usepackage{bm}% bold math
\usepackage{hyperref}% add hypertext capabilities
\usepackage[mathlines]{lineno}% Enable numbering of text and display math
\usepackage{physics}
\usepackage{xcolor}% text color
\usepackage{enumitem} 
\usepackage{accents}
\newcommand{\change}[1]{{\color{black}{#1}}}%highlight changes
\newcommand{\newparagraph}[1]{{\color{black}{#1}}}%highlight new paragraph

\begin{document}

\title{Interplay Between Disorder and Collective Coherent Response: Superradiance and Spectral Motional Narrowing in the Time Domain}% Force line breaks with \\

\author{Hsing-Ta Chen}
\affiliation{Department of Chemistry, University of Pennsylvania, 231 South 34th Street, Philadelphia, Pennsylvania 19104, United States}
\author{Zeyu Zhou}
\affiliation{Department of Chemistry, University of Pennsylvania, 231 South 34th Street, Philadelphia, Pennsylvania 19104, United States}
\author{Maxim Sukharev}
\affiliation{Department of Physics, Arizona State University, Tempe 85287 AZ, USA}
\affiliation{College of Integrative Sciences and Arts, Arizona State University, Mesa 85212, AZ, USA}
\author{Joseph E. Subotnik}
\affiliation{Department of Chemistry, University of Pennsylvania, 231 South 34th Street, Philadelphia, Pennsylvania 19104, United States}
\author{Abraham Nitzan}
\affiliation{Department of Chemistry, University of Pennsylvania, 231 South 34th Street, Philadelphia, Pennsylvania 19104, United States}
\date{\today}

% \begin{document}
\begin{abstract}
The interplay between static and dynamic disorder and collective optical response in molecular ensembles is an important characteristic of nanoplasmonic and nanophotonic molecular systems.
Here we investigate the cooperative superradiant response of a molecular ensemble of quantum emitters under the influence of environmental disorder, including inhomogeneous broadening (as induced by static random distribution of the molecular transition frequencies) and motional narrowing (as induced by stochastic modulation of these excitation energies).
The effect of inhomogeneous broadening is to destroy the coherence of the collective molecular excitation and suppress superradiant emission.
However, fast stochastic modulation of the molecular excitation energy can effectively restore the coherence of the quantum emitters and lead to a recovery of superradiant emission, which is an unexpected manifestation of motional narrowing.
For a light scattering process as induced by an off-resonant incident pulse, stochastic modulation leads to inelastic fluorescence emission at the average excitation energy at long times and suggests that dynamic disorder effects can actually lead to collective excitation of the molecular ensemble.
\end{abstract}
\maketitle

\section{Introduction}
Cooperative light-matter interactions have been observed in a variety of physical systems and utilized for many applications in quantum information processing and cavity polariton chemistry\cite{asenjo-garcia_exponential_2017,asenjo-garcia_optical_2019,thomas_tilting_2019,galego_cavity-induced_2015}.
The essence of such interactions is that many molecules and materials interact together with a common optical field, such as a cavity photon mode or a continuum of radiation fields, leading to a collective optical response that is different from the response of a set of independent molecular emitters. 
In particular, the superradiance phenomenon\cite{dicke_coherence_1954} describes how many quantum emitters can radiate collectively at an enhanced rate that is faster than their individual spontaneous emission rate.
The collective nature of superradiant emission is manifested by the enhanced emission rate that increases with the number of emitters involved in the cooperative light-matter interactions, as observed experimentally in cold atoms in an optical cavity or waveguides\cite{guerin_subradiance_2016,wellnitz_collective_2020,norcia_superradiance_2016,ferioli_laser-driven_2021,bromley_collective_2016,goban_superradiance_2015}, molecular aggregates\cite{spano_superradiance_1989,de_boer_dephasing-induced_1990,fidder_superradiant_1990,gomez-castano_energy_2019,spano_excitonphonon_2020}, nitrogen vacancies in nanocrystals\cite{bradac_room-temperature_2017}, and lead halide perovskite\cite{raino_superfluorescence_2018,mattiotti_thermal_2020}.

% non-Hermitian Hamiltonian
From a theoretical perspective, the simplest model for describing superradiant emission is to propagate open quantum dynamics where the molecular Hamiltonian $\hat{H}_M$ is augmented a non-Hermitian coupling term $\hat{H}_\text{eff}=\hat{H}_M-i\frac{\Gamma}{2}\hat{Q}$. \cite{sokolov_interfering_1997,auerbach_doorway_2007,auerbach_super-radiant_2011} 
Here, $\hat{H}_M$ describes $N$ (ideally non-interacting) subsystems, and  the simplest choice for the non-Hermitian operator $-i\frac{\Gamma}{2}\hat{Q}$ is $\bra{x_i}\hat{Q}\ket{x_j}=1$ for all $x_i,x_j$ molecular excited states. In this expression, $\Gamma$ is the single molecule spontaneous emission rate.
The non-Hermitian operator above accounts for the \emph{overall} effects of cooperative light-matter interactions (assuming that every molecule equally couples to the radiation fields). 
The effective Hamiltonian has one complex-valued eigenvalue and $N-1$ real-valued eigenvalues.
The eigenstate corresponding to the complex-valued eigenvalue, referred to as the superradiant state, is a coherent superposition of \emph{all} the molecular excitations.
This superradiant state is bright in the sense that the coherent molecular excitation decays and emits radiation at the superradiance rate $N\Gamma$.
In contrast, other eigenstates corresponding to the real eigenvalues, so-called dark states, do not decay as a function of time. 

%Beyond Effective Non-Hermitian Hamiltonian
It should be noted that the simple model above has a few limitations. 
First, the model does not use an explicit description of photonic states, which will be important for the present manuscript. 
To overcome this limitation, below we will explicitly model a bath of photon modes all coupled to the set of emitters (rather than use $\Gamma\hat{Q}$). This present approach will allow us to address the distribution of emitted photons while still being computationally affordable.
A second limitation of the model above is the assumption that all molecules see the same photon manifold (embedded in the form postulated for the matrix $\hat{Q}$); this assumption should hold if the molecular system is small relative to the relevant wavelength and if one can disregard orientational disorder, \change{which may occur in realistic systems, such as fluorescent dyes\cite{gomez-castano_energy_2019}, ordered superlattices of perovskite\cite{raino_superfluorescence_2018,zhou_cooperative_2020}, and closely packing quantum dots\cite{tahara_collective_2021}}. 
Now, in principle, for emitters placed within a lattice or an atomic array\cite{mattiotti_thermal_2020,asenjo-garcia_optical_2019}, the corresponding molecule-photon couplings should be a function of the relative position of the emitters.
This spatial dependence can modify the eigenstates and eigenvalues of the effective Hamiltonian, so that the dark states acquire weak decay rates and become the so-called subradiant states\cite{auerbach_super-radiant_2011,zhang_theory_2019,zhang_subradiant_2020,bienaime_controlled_2012}. 
In turn, the interplay between the superradiant and subradiant states is responsible for the biexponential decay profiles that are observed in molecular aggregates and quantum dot arrays\cite{lim_exciton_2004,sergeev_tailoring_2020}, whereupon the collective emission changes from a fast decay at short times to a much slower decay for the long-time dynamics.
Nevertheless, even though below we will discuss the transition from superradiance to subradiance, we will not concern ourselves with how spatial placing effects the relaxation operator.  
%We note that, for small molecular aggregates, such spatial dependence can be disregarded
\change{As a third limitation of the model, we consider molecular interactions mediated only by optical photon exchange and disregard electrostatic interactions, which usually cannot be ignored in realistic systems.
However, Ref.~\cite{gomez-castano_energy_2019} shows that some collective interference phenomena in excitation energy transfer between molecular aggregates can be captured by simple semiclassical electrodynamic simulations without accounting for intermolecular coupling.
Thus, this assumption may have the potential to hold in some systems. }
Future work can certainly address this shortcoming of our model.

In what follows below, we will demonstrate that subradiant features emerge naturally in very simple simulations due to disorder effects as induced by interaction with the environment.
%disorder effects in the literature
Historically, most published studies on disorder in molecular ensembles have considered static disorder \citep{celardo_interplay_2013,celardo_cooperative_2014,giusteri_non-hermitian_2015,biella_subradiant_2013}, where environmental processes are assumed to be much slower than the timescale of radiative relaxation. 
As such, each molecule experiences a slightly different local environment leading to a random dipolar orientation and fluctuating electronic transition frequency in the model.
Fewer studies in the literature have considered dynamic disorder, where disorder has a timescale comparable or faster than molecular emission as induced by the thermal motion of the environment (which is sometimes just treated as a phenomenological molecular dephasing rate\cite{temnov_superradiance_2005}). 
Formally, the definition of dynamic disorder (for an array of emitters) is to consider either the stochastic modulation of the dipolar orientation or the excitation energy of each molecule, which altogether can lead to many interesting phenomena.
% recent experiments report that an off-resonance light source can lead to slow, subradiant fluorescence emissions from a disorder ensemble of emitters
For instance, a slow, subradiant fluorescence signal at long times can be observed in a cold atom cloud excited by an off-resonance laser pulse\cite{guerin_subradiance_2016,weiss_robustness_2019,rui_subradiant_2020}.
Moreover, recent experiments show that motional narrowing\footnote{\change{The term ``motional narrowing'' is widely used in the context of nuclear magnetic resonance---as the atoms move in an inhomogeneous medium, the norm of the fluctuations of the time-averaged magnetic field experienced by the atoms is smaller than the standard deviation of the static magnetic field when the atoms are stationary. As a consequence, the absorption linewidth becomes narrower when the motion of the atoms is considered.
In Kubo's stochastic modulation model, the effect of the atomic motion is mathematically modeled as the time-dependent fluctuation of the atomic transition energy.
Following the same path, our model considers a set of molecules interacting with the radiation field cooperatively and experiencing randomly fluctuating environmental configuration as induced by the thermal motion.}} as induced by dynamic disorder\cite{w_anderson_mathematical_1954,kubo_stochastic_1969} can be used to entangle quantum states and restore coherence for quantum emitters\cite{empedocles_quantum-confined_1997,berthelot_unconventional_2006,pont_restoring_2021}.
\change{At this point, one may wonder: when the temperature increases, can the effect of fast modulation be observed even in the presence of other temperature-dependent effects?
In fact, absorption spectrum narrowing has been observed in some molecular systems when structural transitions (which arise with increasing temperature) activate fast local motion of cation or anion molecules; such systems include plastic crystals\cite{salgado-beceiro_motional_2018} and hybrid organic-inorganic perovskites\cite{koda_organic_2022}.}
Interestingly, however, to our knowledge, the effects of such dynamic disorder on collective emission have not been explored beyond a phenomenological dephasing approximation.

With this background, in the present paper, we investigate the cooperative emission of a molecular ensemble under the influence of environmental dynamic disorder characterized stochastic modulation timescales ranging from fast (relative to all molecular timescales) to slow modulation down to the static limit. 
We focus first on the case in which the molecules are prepared initially in the coherent excited state, and we analyze the crossover between superradiance and subradiance under static and dynamic disorder. 
The second focus of the paper is the off-resonant light scattering process of a disordered molecular ensemble.
Such scattering is necessarily elastic when the molecular target is static, however in the presence of fast dynamic disorder (pure dephasing), inelastic fluorescence emission accompanies the elastic scattering signal. 
Here we discuss the onset of this phenomena and how it reflects the collective nature of the many-molecule response.
The outline of the paper is as follow:
In section \ref{model}, we formulate a model for collective emission that treats the radiation fields explicitly and introduce static and dynamic disorder within the model.
In section \ref{sec:superradiance}, we investigate the superradiant emission from the coherernt state under the influence of dynamic disorder and discuss how the effect of motional narrowing can be manifested in the time domain.
In section \ref{sec:pulse}, we focus on the optical response of a disordered molecular ensemble interacting with an off-resonant light pulse and elucidate the collective feature of the inelastic fluorescence signals. 
We conclude in section \ref{sec:conclusion}.

\section{Model\label{model}}
\subsection{Model Hamiltonian}
Processes involving collective light-matter interactions can be modeled by an ensemble of $N$ quantum emitters coupled to a shared continuum of photon states using the machinery of quantum electrodynamics.
The total Hamiltonian takes the form of $\hat{H}=\hat{H}_{M}+\hat{H}_{R}+\hat{V}_{MR}$ where $\hat{H}_{M}$ is the Hamiltonian of the molecular subsystem, $\hat{H}_{R}$ is the quantized photon Hamiltonian, and $\hat{V}_{MR}$ describes the molecule-radiation coupling.
The molecular subsystem is composed of $N$ two-level systems where the $j$-th molecule has the ground state $\ket{g_j}$ , the excited state $\ket{x_j}$, and the electronic transition frequency $\omega_j$ . 
In the span of the total ground state $\ket{G}=\prod_{k=1}^{N}\ket{g_j}$ and the single excitation Fock states $\ket{X_j}=\ket{x_j}\prod_{k\neq j}^{N}\ket{g_k}$  (only the $j$-th molecule is excited), the total Hamiltonian of the molecular subsystem takes the form
\begin{equation}
\hat{H}_{M}=E_G\ket{G}\bra{G}+\sum_{j=1}^{N}E_j\ket{X_j}\bra{X_j}.
\end{equation}
Here the single molecular excitation energy is given by $E_j-E_G=\hbar\omega_j$ where $E_G$ denotes the ground state energy and the intermolecular coupling is disregarded. 
Within the single excitation subspace, we model the radiation fields as a set of single photon states  $\{\ket{\alpha}\}$  with frequencies $\{\omega_\alpha\}$ and the single photon Hamiltonian is 
\begin{equation}
% \hat{H}_{R}=\sum_{\alpha}(\hbar\omega_\alpha-i\frac{\eta}{2})\ket{\alpha}\bra{\alpha}.
\hat{H}_{R}=\sum_{\alpha}\hbar\omega_\alpha\ket{\alpha}\bra{\alpha}.
\end{equation}
The molecule-radiation coupling takes the form of 
\begin{equation}
\hat{V}_{MR}=\sum_{j,\alpha}V_{j,\alpha}\left(\ket{X_j}\bra{\alpha}+\ket{\alpha}\bra{X_j}\right)
\end{equation}
where $V_{j,\alpha}$ denotes the single-molecule coupling strength of the $j$-th molecule to the photon mode $\omega_\alpha$.
% Note that this simplified model relies on several approximations (see appendix).
As a final note, we emphasize that this model for collective emission involves single excitation states as coupled to a shared continuum of photon states\cite{svidzinsky_cooperative_2010}, rather than the original Dicke superradiance problem where all molecules are initially excited (which would require a set of quantum state with $N$ excitations rather than single excitations).

% introduce FGR rate and superradiance state 
In the absence of disorder, we assume that the emitters are identical, i.e. $\omega_j=\omega_x$ is the same for all molecules, and make several further assumptions as follows.
First, we assume the system is small relative to the radiation wavelength (the long wavelength approximation) and disregard orientational disorder so that $V_{j,\alpha}=v_\alpha$ for all $j$.
Second, we employ the wide-band approximation (i.e. $v_\alpha$ is a constant for all $\alpha$), so that each single emitter decays and emits photons to the radiation continuum at the spontaneous emission rate $\Gamma=2\pi\sum_\alpha{|v_\alpha|^2\delta(\omega_x-\omega_\alpha)}$, as one can derive by the Wigner-Weisskopf theory. 
Under these assumptions, the superradiant state of the molecular subsystem is the fully symmetric superposition of \emph{all} the single excitation states, $\ket{S}=\frac{1}{\sqrt{N}}\sum_{j=1}^{N}\ket{X_j}$.
When the molecular ensemble is initially prepared in the superradiant state, the total excitation population should decay and emit photons at an enhanced rate $N\Gamma$. 
Note that the enhancement of the emission rate by the number of emitters $N$ is a signature of the collectivity of the superradiant emission.
%  By projecting the overall effects of the photon field ($\hat{H}_{R}+\hat{V}_{MR}$)\cite{sokolov_interfering_1997,auerbach_super-radiant_2011}, the total Hamiltonian can be reduced to an effective non-Hermitian Hamiltonian $\tilde{H}=\hat{H}_{M}+\hat{V}_{ext}(t)-\frac{i}{2}\Gamma\hat{Q}$ 
% Here, $\hat{Q}=\sum_{jk}\ket{X_j}\bra{X_k}$ and $\Gamma=2\pi\sum_\alpha{|v_\alpha|^2\delta(E_x-\hbar\omega_\alpha)}$ is the Fermi's golden rule rate for single molecule spontaneous emission (we use the wide-band approximation).

The dynamics of the total system are governed by the time-dependent Schrodinger equation $\frac{d}{dt}\ket{\psi(t)}=-i\hat{H}(t)\ket{\psi(t)}$ where the total wavefunction is written in the single excitation subspace$\ket{\psi(t)}=C_0(t)\ket{G}+\sum_{j}{C_j(t)\ket{X_j}}+\sum_{\alpha}{C_{\alpha}(t)\ket{\alpha}}$.
(Throughout this paper, we set $\hbar =1$.) 
The dynamics in the excited molecules subspace is thus given by
\begin{equation}\label{eq:electronic_EOM}
    \frac{dC_j}{dt}=-i\omega_jC_j-i\sum_\alpha V_{j,\alpha}C_\alpha.
\end{equation}
In order to avoid the photonic back-action towards the molecular subsystem, we introduce a damping parameter $\eta$ in the photon modes, i.e.
\begin{equation}\label{eq:photon_EOM}
\frac{dC_\alpha}{dt}=-i\omega_\alpha C_\alpha-i\sum_jV^*_{j,\alpha}C_j-\frac{\eta}{2}C_\alpha.
\end{equation}
The results reported below do not depend on the choice of $\eta$ provided that $\eta$ exceeds the spacing between the energies $\hbar\omega_\alpha$. 
The molecular excitation population (the total probability of finding the excitation in the molecular subsystem) can be calculated by  
\begin{equation}\label{eq:Pop}
    P(t)=\left< \sum_{j=1}^{N}|C_j(t)|^{2}\right>,
\end{equation} 
while the cumulative emission at frequency $\omega_\alpha$ is given by %photon state population distribution
\begin{equation}\label{eq:Icum}
    I(\omega_\alpha,t)=\left<|C_\alpha(t)|^2+\eta\int_0^t{dt'}|C_\alpha(t')|^2\right>.
\end{equation}
\change{Here $I(\omega_\alpha,t)$ includes the instantaneous population of the photon mode $|C_\alpha(t)|^2$ as well as the damped population.} % For spectroscopic measurement for a given measurement time $T_\text{obs}$, we are interested in the time-resolved emission spectrum as given by 
% \begin{equation}\label{eq:Iobs}
%     I_\text{obs}(\omega_\alpha,t)=\frac{1}{T_\text{obs}}\int_{t}^{t+T_\text{obs}}dt' I(\omega_\alpha,t')\xrightarrow{T_\text{obs}\rightarrow0}\left<|C_\alpha(t)|^2\right>.
% \end{equation}
In general, for the parameters chosen in this paper, propagating the molecular subsystem wavefunction with the protocol above leads to results that are equivalent to those obtained from an effective non-Hermitian Hamiltonian $\hat{H}_\mathrm{eff}=\hat{H}_M-i\frac{\Gamma}{2}\hat{Q}$.
%yields almost the same molecular excitation population dynamics, but cannot describe properties of the emitted photon (such as the emission spectrum).

\subsection{Dynamic and static disorder}
We now consider disorder effects as induced by the interaction of the molecular ensemble with the environment. 
Depending on the timescale of the environmental process relative to the molecular emission, the molecular ensemble can experience two types of disorder.

\subsubsection{Static disorder}
For a process that is much slower than molecular emission, the local environment can be considered time-independent and the inhomogeneity of the environment leads to a statistical distribution of the electronic transition frequency. 
In practice, static disorder is usually modeled by including a random component in the electronic transition frequency
\begin{equation}
\tilde{\omega}_j=\omega_x+\delta{\omega_j}.\label{eq:static_disorder}
\end{equation}
Here $\delta{\omega_j}$ is a random variable satisfying $\langle\delta{\omega_j}\delta{\omega_k}\rangle=\sigma^2\delta_{jk}$ where $\delta_{jk}$ is a Kronecker delta function. We take $\{\delta{\omega_j}\}$ to be a Gaussian random variable chosen according to the probability distribution $\mathrm{Prob}(\delta{\omega_j})=\frac{1}{\sqrt{2\pi\sigma^{2}}}e^{-\delta{\omega_j}^2/2\sigma^2}$ and the width of the distribution $\sigma$ characterizes the disorder amplitude. 
The random variable satisfies $\left<\delta{\omega_j}\right>=0$ where $\left<\cdots \right>$ denotes averaging over realizations, so the average electronic transition frequency is $\left<\tilde{\omega}_j\right>=\omega_x$.
Similarly, observables are ensemble averages over different realizations of Eq.~\eqref{eq:static_disorder}.

\subsubsection{Dynamic disorder}
When an environmental process is faster than or comparable with molecular emission, each molecule experiences a time-dependent, randomly fluctuating environmental configuration as induced by the thermal motion.
Such a dynamic disorder can be modeled by including a time-dependent modulation to the  electronic transition frequency
\begin{equation}
\omega_j(t)=\omega_x+\Omega_{j}(t)\label{eq:dynamic_disorder}
\end{equation}
where $\Omega_{j}(t)$ is a stochastic variable as in Kubo's stochastic modulation model\cite{kubo_stochastic_1969}.
Here we choose $\Omega_{j}(t)$ to be a Gaussian stochastic variable satisfying $\langle\Omega_{j}(t)\rangle=0$ for all $j$ and the correlation function is  
\begin{equation}
    \langle\Omega_{j}(t_{1})\Omega_{k}(t_{2})\rangle=\delta_{jk}\sigma^{2}e^{-|t_{1}-t_{2}|/\tau_{c}}\label{eq:dynamic_correlation}
\end{equation} 
where $\delta_{jk}$ is a Kronecker delta function.
Note that Eq.~\eqref{eq:static_disorder} is the static limit ($\tau_c\rightarrow\infty$) of a general stochastic process in which $\delta\omega_j$ varies in time as a stochastic variable.

The Gaussian stochastic frequency defined by Eqs.~\eqref{eq:dynamic_disorder} and \eqref{eq:dynamic_correlation} is characterized by two parameters: (i) the disorder amplitude $\sigma$ indicates the strength of the stochastic modulation, and (ii) the correlation time  $\tau_{c}$ estimates the timescale for how rapidly $\Omega_{j}(t)$ changes.
Namely, the smaller  $\tau_{c}$ is, the faster  $\Omega_{j}(t)$ modulates the electronic transition frequency.
In numerical practice, the Gaussian stochastic variable for a discrete time series $t_0,t_1,\cdots,t_n$ with $t_i=idt$ can be generated from a Markovian process that the probability distribution of $\Omega_{j}(t_i)$ depends only on the immediately previous value $\Omega_{j}(t_{i-1})$ (see Appendix \ref{sec:generate}).

According to Kubo's lineshape theory\cite{kubo_stochastic_1969}, one defines $1/\sigma\tau_{c}$ as the modulation rate.
In the slow modulation limit ($1/\sigma\tau_{c}\ll1$), the Gaussian stochastic variable $\Omega_{j}(t)$ becomes effectively time-independent and one recovers the static disorder case corresponding to a Gaussian probability distribution with disorder amplitude $\sigma$ as in Eq.~\eqref{eq:static_disorder}.
In the fast modulation limit ($1/\sigma\tau_{c}\gg1$), the time correlation function becomes $\langle\Omega_{j}(t_{1})\Omega_{k}(t_{2})\rangle\rightarrow\delta_{jk}\sigma^{2}\times2\tau_{c}\delta(|t_{1}-t_{2}|)$ where $\delta(|t_{1}-t_{2}|)$ is a Dirac delta function.
Such a fast random energy modulation implies that, within the timescale of molecular emission, each molecule can almost experience all accessible configurations of its local environment. In this limit, the overall effect of modulating the transition frequency stochastically is equivalent to including an effective molecular dephasing at the dephasing rate $\gamma=\sigma^{2}\tau_{c}$ (which ends up being the width of the lineshape function in Kubo's theory).\cite{kubo_stochastic_1969}  
% Therefore, one can implement this dephasing mechanism by choosing $\Omega_{j}(t)$ as a Gaussian random variable with $\langle\Omega_{j}(t)^2\rangle=\gamma/2dt$ at each time step.

\begin{figure}
\includegraphics{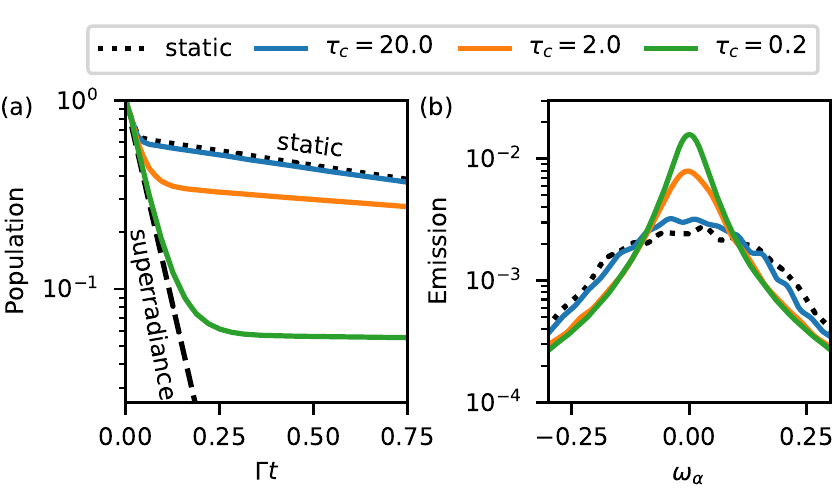} 
\caption{(a) Molecular excitation survival probability and (b) the corresponding cumulative emission spectrum at $\Gamma t=2$ are plotted for $\tau_c=20,2,0.2$. 
% Note that this simulation has not yet reach $P=0$ at $\Gamma t=2$, so the integrals of the emission over $\omega_\alpha$ are not the same for different $\tau_c$ in (b). 
The initial state of the molecular ensemble is the superradiant state $\ket{S}$, Eq.~\eqref{eq:initial_psi}, and the disorder amplitude is $\sigma=0.1$.
Without disorder, the excitation population decays at the superradiance rate $N\Gamma$ (dashed lines).
For a molecular ensemble with static disorder (black dotted lines in (a)), the excitation population shows a biexponential decay ($e^{-N\Gamma t}$ for the superradiance at short times and $e^{-\Gamma' t}$ for the subradiance in the long time), and the corresponding emission spectrum is a Gaussian distribution (black dash lines in (b)).
For the dynamic disorder cases (solid lines), as $\tau_c$ decreases, we find that the shape of the emission spectrum becomes narrower and turns into a Lorentzian distribution (as observed in (b)). 
% The spectrum with the most motional narrowing in (b) corresponds to the population dynamics whereby the molecular excitation decays superradiantly for the longest period of time in (a).
Coincidentally, the superradiant component in the time-resolved relaxation seen in (a) becomes more dominant, signifying recovery of coherent superradiant emission.}\label{fig1}
\end{figure}

\begin{figure}
\includegraphics{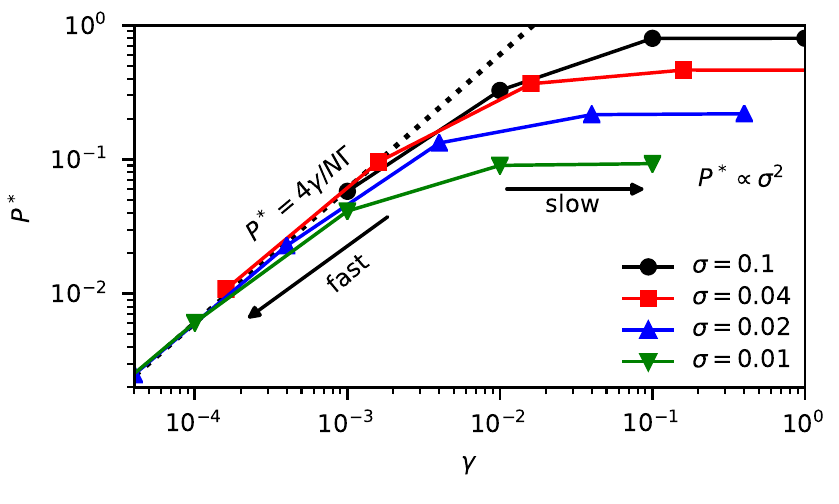} 
\caption{The critical population $P^*$ is plotted as a function of the dephasing rate $\gamma=\sigma^2\tau_c$. 
% The black solid line is calculated using the dephasing method. 
The colored lines are obtained by implementing Gaussian stochastic modulation and, for each line, we fix $\sigma$ and plot $P^*$ with varying $\tau_c$.
In the fast modulation limit (small $\tau_c$, left), $P^*$ converges to the black dotted line \newparagraph{as predicted by Eq.~\eqref{eq:Pstar_estimate}} and linearly depends on $\gamma$.
In the slow modulation limit (large $\tau_c$, right), $P^*\propto\sigma^2$ does not depend on $\tau_c$ and approaches the corresponding static disorder results.
% \newparagraph{The lower panel is the approximate $P^*$ calculated using Eq.~\eqref{eq:t_critical}. In the fast modulation limit, the approximate $P^*$ scales linearly with $\gamma$ and agrees with the numerical results. In the static limit, the approximate $P^*$ does not depends on $\gamma$, but is over-estimated when $\sigma$ is large. Note that, for $\sigma=0.1$, the perturbative approximation is not valid leading to unphysical results ($P^*>1$). }
}\label{fig2}
\end{figure}

\newparagraph{
\subsection{Perturbative analysis of disorder effects}
Having introduced the different types of disorder, we are now ready to analyze disorder effects on the excitation population dynamics based on the effective non-Hermitian Hamiltonian ($\hat{H}_\mathrm{eff}=\hat{H}_M-i\frac{\Gamma}{2}\hat{Q}$) where $\hat{Q}=\sum_{jk}\left|X_{j}\right\rangle \left\langle X_{k}\right|$.
We assume that the molecular ensemble is initially prepared in the fully symmetric single excitation state \begin{equation}\label{eq:initial_psi}
\ket{\psi(0)}=\ket{S}=\frac{1}{N}\sum_{j=1}^{N}\ket{X_j}.
\end{equation}
Within our model, such a state can be formed by excitation from the ground state using a short broadband excitation (approximately a $\delta$-function pulse).
Note that, without disorder, the excitation population decays at the superradiance rate $P(t)=e^{-N\Gamma t}$ (see the black dashed line in Fig.~\ref{fig1}(a)).
\change{For the case of static disorder, one can average the behavior of the eigenstates of $\hat{H}_\mathrm{eff}$ with complex-valued eigenvalues (where the imaginary part of the eigenvalue corresponds to the decay rate of the eigenstate) and explain the biexponential decay of the excitation population---a superradiant decay for short times and then evolves to a subradiant decay for long times.\cite{celardo_cooperative_2014} However, such eigenvalue analysis cannot be easily done for the case of dynamic disorder, which is discussed below.}

To analyze the population relaxation under dynamic disorder, we employ time-dependent perturbation theory\cite{fetter_quantum_2003,nitzan_chemical_2006} and divide the effective non-Hermitian Hamiltonian into $\hat{H}_\mathrm{eff}=\hat{H}_0+\hat{V}(t)$, where the fluctuations of the electronic transition frequency are treated as a time-dependent perturbation $\hat{V}(t)=\sum_{j}\Omega_{j}(t)|X_{j}\rangle\langle X_{j}|$ and the unperturbed Hamiltonian is $\hat{H}_{0}=\sum_{j}\omega_x\left|X_{j}\right\rangle \left\langle X_{j}\right|-i\frac{\Gamma}{2}\sum_{jk}\left|X_{j}\right\rangle \left\langle X_{k}\right|$.
With this perturbation, the propagator of the electronic wavefunction can be expanded in terms of multi-time integrals of $\hat{V}(t)$ (see Eq.~\eqref{eq:propagator_all}).
Next we gather the zeroth and first order terms of the propagator and approximate the excitation population dynamics as (see Appendix.~\ref{sec:TDPT} for more detail)
\begin{equation}
    P(t)= P_s^{(0)}(t) + P_s^{(1)}(t) + P_d^{(1)}(t). 
\end{equation}
Here, 
\begin{equation}
    P_s^{(0)}(t) = e^{-N\Gamma t}
\end{equation}
is from the zeroth order term and is responsible for the collective superradiant emission at short times;
\begin{equation}
  P_s^{(1)}(t)=\frac{2}{N}e^{-N\Gamma t}\gamma\left(t+\tau_{c}e^{-t/\tau_{c}}-\tau_{c}\right).
\end{equation}
is relatively small and contributes to only the transient dynamics (i.e. $P_s^{(1)}(0)=0$ and $P_s^{(1)}(t\rightarrow\infty)=0$);
\begin{equation}
    P_d^{(1)}(t)=\frac{2(N-1)\sigma^{2}}{N^{2}\Gamma}\left(\frac{1}{q}e^{-N\Gamma t}+\frac{1}{r}-\frac{N\Gamma}{rq}e^{-rt}\right)
\end{equation}
where $r=\frac{N\Gamma}{2}+\frac{1}{\tau_{c}}$ and $q=\frac{N\Gamma}{2}-\frac{1}{\tau_{c}}$. 
We find that $P_d^{(1)}(t)$ emerges from zero at $t=0$ and dominates the dynamics at long times.
Therefore, we can now estimate the critical $t^*$ at which the dynamics turns from $P_s^{(0)}(t)$ to $P_d^{(1)}(t)$ by letting $P_s^{(0)}(t^*)=P_d^{(1)}(t^*)$, i.e. 
\begin{equation}\label{eq:t_critical}
    e^{-N\Gamma t^{*}}=\frac{2(N-1)\sigma^{2}}{N^{2}\Gamma}\left(\frac{1}{q}e^{-N\Gamma t^*}+\frac{1}{r}-\frac{N\Gamma}{rq}e^{-rt^*}\right).
\end{equation}
The estimated critical population at this time is $P^*=2P_s^{(0)}(t^*)$.
In the fast modulation limit ($\tau_{c}\rightarrow0$ or $\tau_{c}\ll\frac{2}{N\Gamma}$), since $r\rightarrow1/\tau_c$, $q\rightarrow-1/\tau_c$, Eq.~\eqref{eq:t_critical} leads to (assuming $N$ is large)
\begin{equation}\label{eq:Pstar_estimate}
    P^*|_{\tau_c\rightarrow0}= \frac{N-1}{N^{2}}\frac{4\gamma}{\Gamma}\approx\frac{4\gamma}{N\Gamma}.
\end{equation}
\begin{equation}
    t^*|_{\tau_c\rightarrow0}\approx\frac{1}{N\Gamma}\ln\left(\frac{N\Gamma}{2\gamma }\right).
\end{equation}
Thus, for a fixed disorder amplitude $\sigma$, the estimated time span for the collective emission becomes longer (i.e. $t^*$ increases) when the stochastic modulation becomes faster (i.e. $\gamma$ or $\tau_c$ decrease).

% In the static limit ($\tau_{c}\rightarrow\infty$, or $\tau_{c}\gg\frac{2}{N\Gamma}$), we have $r=q\rightarrow N\Gamma/2$ and Eq.~\eqref{eq:t_critical} becomes
% \begin{equation}
%     P^*|_{\tau_c\rightarrow\infty}= \frac{N-1}{N^{2}}\frac{8\sigma^{2}}{N\Gamma^{2}}\approx\frac{8\sigma^{2}}{N^{2}\Gamma^{2}}
% \end{equation}

However, we notice that $P_d^{(1)}(t)$ does not decay to zero at long times (see Eq.~\eqref{eq:Pd1_infty}) and cannot capture the subradiant decay qualitatively.   
Effectively, $P_d^{(1)}(t)$ arises from the first order terms in which the unperturbed Hamiltonian $H_0$ is perturbed by the time-dependent fluctuation $V(t_1)$ just once at $t_1$ and the fully symmetric state $|S\rangle$ is not completely bright for the perturbed Hamiltonian $H_0+V(t_1)$.
As a result, after $t_1$, the dark part of the electronic state does not decay as we consider only up to the first order terms.
That being said, Fig.\ref{fig:Kubo} in Appendix~\ref{sec:TDPT} shows that, in the small dephasing rate limit ($\gamma\ll\Gamma$), this perturbative approximation can almost accurately capture the population relaxation within the time span when the transition occurs, leading to a quantitative prediction of $P^*$. The actual behavior of $P^*$ is analyzed in Fig.~\ref{fig2} numerically.
}

\section{The Effect of Disorder on Superradiant Emission}\label{sec:superradiance}
% Having introduced the different types of disorder, we are now ready to investigate the dynamical interplay of the cooperative emission with static and dynamic fluctuations.
% We assume that the molecular ensemble is initially prepared in the fully symmetric single excitation state \begin{equation}\label{eq:initial_psi}
% \ket{\psi(0)}=\ket{S}=\frac{1}{N}\sum_{j=1}^{N}\ket{X_j}.
% \end{equation}
% Within our model, such a state can be formed by excitation from the ground state using a short broadband excitation (approximately a $\delta$-function pulse).
% Note that, without disorder, the excitation population decays at the superradiance rate $P(t)=P(0)e^{-N\Gamma t}$ (see the black dashed line in Fig.~\ref{fig1}(a)).

\change{With this analytical intuition in mind, we will now numerically investigate the dynamical interplay of the cooperative emission with static and dynamic fluctuations.}
In the calculation reported below, we use a molecular ensemble of $N=20$ emitters and choose the average excitation energy $\omega_x=1$ as the unit of energy.
The continuum of photon states is explicitly described by a set of single photon states with frequency $\omega_\alpha=d\omega(\alpha-M/2)$ for $\alpha=0,\ldots,M$, with interlevel spacing $d\omega=2\times10^{-3}$, a bandwidth determined by $M=600$, and a damping parameter (see Eq.~\eqref{eq:photon_EOM})is chosen to be $\eta=0.01$.
The molecule-radiation coupling is uniform $V_{j,\alpha}=10^{-3}$ and consequently the single-molecule spontaneous emission rate is $\Gamma=\pi\times10^{-3}$. 
For the ensemble average $\left<\cdots\right>$ in the following results, we average $256$ realizations \change{(which, we found, is sufficient to achieve $\sqrt{\langle(P-\langle{P}\rangle)^2\rangle}/\langle{P}\rangle<0.1$)}.
The results reported below do not depend on the choice of bandwidth $Md\omega$ or $\eta$, provided that $\eta>d\omega$ and $Md\omega>N\Gamma$.

\subsection{Motional narrowing manifested in the frequency domain and in the time domain}

Fig.~\ref{fig1} shows the excitation population dynamics $P(t)$ and the corresponding cumulative emission spectrum $I(\omega_\alpha,t=2/\Gamma)$ for different disorder profiles.
In general, in the presence of disorder (either static or dynamic), the excitation population shows a biexponential decay, rather than a single superradiant decay.
Specifically, $P(t)$ decays at the superradiant rate $N\Gamma$ (along the black dashed line) for short times and then evolves to follow a subradiant decay rate $\Gamma'<\Gamma$ at long times.
In the presence of dynamic disorder, we find that, as expected, if the correlation time is long ($\tau_c=20$), the dynamics of the excitation population almost recovers the dynamics of the static disorder case (the black dotted line). 
More importantly, for a fixed disorder amplitude $\sigma$, as the correlation time $\tau_c$ becomes shorter (i.e. $\omega_j(t)$ modulates more rapidly), more excitation population decay occurs at the superradiant rate before the decay becomes subradiant.
The increasing fraction of the superradiant decay in the fast modulation limit (versus the subradiant decay in the long time limit) implies that 
%including dynamic disorder actually preserves the coherence of the initial superradiant state which is itself destroyed by static disorder (i.e. environmental inhomogeneity).
the coherence of the superradiant state, which is quickly destroyed by static disorder, is preserved or recovered when the disorder modulation becomes faster even as the disorder amplitude remains constant.
We thus observe that fast stochastic modulation, that is known to lead from Gaussian lineshape associated with static disorder (seen for $\tau_c=20$) to a motionally narrowed Lorentzian lineshape ($\tau_c=0.2$), is also expressed in the time domain as preservation of the coherent superradiant decay. 
It appears that fast stochastic modulation results in recovery of the collective behavior, that is effective elimination of the decoherence caused by static disorder.

\subsection{Convergence in the fast modulation limit}
% With this picture in mind, one may wonder: if we keep decreasing $\tau_c\rightarrow0$, how does the collective behavior depend on $\sigma$ and $\tau_c$?  
The correlation between the dynamic disorder correlation time $\tau_c$ and the persistence of the superradiant emission, together with the analysis made above, suggests that this behavior is a manifestation of the motional narrowing phenomenon.
To further quantify this observation, we have fitted the population dynamics $P(t)$ (from Fig.~\ref{fig1}) to a bi-exponential functional form 
\begin{equation}
P(t)\approx P(0)f(t-t^*)e^{-N\Gamma t}+(1-f(t-t^*))ae^{-\Gamma' t}. 
\end{equation}
Here $\Gamma'$ is the subradiant rate as obtained by fitting the long-time decay to $ae^{-\Gamma' t}$ and $f(t)=\frac{1}{2}-\frac{1}{\pi}\tan^{-1}(t/b)$ is a smooth step function. 
This biexponential fitting yields the critical time $t^*$ at which the population dynamics changes from a superradiant decay ($e^{-N\Gamma t}$) to a subradiance decay ($e^{-\Gamma' t}$), as well as the population at this time $P^*=P(t^*)$.
% We also find that $t^*\approx\frac{-\log{a}}{N\Gamma-\Gamma'}$ and $P^*$ is not sensitive to the fitting parameter $b$.
The fraction $R_s=\frac{P(0)-P^*}{P(0)}$ quantifies how much of the initially excited population decays at the superradiant rate.

In Fig.~\ref{fig2}, we plot $P^*$ as a function of the dephasing rate $\gamma=\sigma^2\tau_c$ for different disorder amplitudes (since we set $P(0)=1$, $R_s=1-P^* $).
The following observations are noteworthy: 
(i) For a fixed $\sigma$, as $\tau_c$ decreases (faster stochastic modulation), $P^*$ becomes small and $R_s\rightarrow 1$, implying that more of the decay is of the superradiant character.
(ii) In the fast modulation limit ($\gamma\rightarrow0$), $P^*$ for different $\sigma$'s converges and depends linearly on the dephasing rate $P^*\propto\gamma$.
(iii) In the slow modulation limit ($\gamma\rightarrow\infty$, i.e. static disorder), $P^*$ becomes independent of $\tau_c$ and asymptotically approaches different values depending on $\sigma$,  i.e. $P^*\propto\sigma^2$ as $\tau_c\rightarrow\infty$. 
Note that the asymptotic relation ($P^*\propto\sigma^2$) does not hold in the strong disorder limit (when $\sigma$ gets large)---after all, when $\sigma\gg N\Gamma$, the molecular ensemble should behave like a set of independent emitters and the excitation population decays at the spontaneous single molecule emission rate, rather than a bi-exponential decay.
In fact, for this reason, $P^*$ is not really well-defined in the limit $\sigma\rightarrow \infty$.
% This relation in the static disorder limit agrees with the intuition that, if the disorder amplitude $\sigma$ increases, one expect less decay can occur at the superradiance rate (i.e. large $P*$).
\newparagraph{As a final note, we find that, qualitatively, these observations agree with the analytical results as estimated by Eq.~\eqref{eq:t_critical}).
Particularly, for a fixed $\sigma$ in the fast modulation limit (small $\tau_c$), the numerical results approaches $P^*=4\gamma/N\Gamma$ as one expect in Eq.~\eqref{eq:Pstar_estimate}.}

\section{Off-resonant Light Scattering for a Molecular Ensemble}\label{sec:pulse}
%AN: start by reviewing the history of this observation.
The previous section has analyzed the effect of disorder on molecule-radiation interactions under the assumption that all dynamics are initialized in a bright state. 
More generally, one would like to model the decay that arises for a system that is pumped with external light.
For a single molecule in the absence of dephasing, light scattering processes can be described by a model that couples the molecule to an external incoming field; the molecule emits photons into the radiation continuum that can be observed as a scattering signal\cite{tannor_introduction_2006}.
For incoming light that is resonant with the molecular excitation, the pulse can raise the population of a molecular excited state and, following the pulse, the molecule emits fluorescence at the spontaneous emission rate. 
% If the frequency of the incoming light is detuned (off-resonance scattering), the excited state of the molecular system cannot be populated after the pulse is switched off and the observed scattering signal will have the same frequency as the incoming light (so-called elastic scattering).
In contrast, an off-resonant pulse cannot populate the molecular excited state so the molecular response appears only during the pulse. 
In either case, in absence of environmental interactions (here expressed by dynamic disorder), light scattering is elastic.

Let us now turn our attention to such a light scattering process from a disordered ensemble of molecules.
Recent experiments report that illumination of a disordered ensemble of molecules with an off-resonance light source can lead to slow, subradiant fluorescence emission\cite{guerin_subradiance_2016,weiss_robustness_2019}.
For our purposes, the relevant Hamiltonian is $\hat{H}+\hat{V}_\mathrm{ext}(t)$, where $\hat{V}_\mathrm{ext}(t)$ captures how the incoming external field couples the electronic ground state to the excited state:
\begin{equation}
\hat{V}_\mathrm{ext}(t)=\sum_{j=1}^N F_j(t)\left(\ket{G}\bra{X_j}+\ket{X_j}\bra{G}\right).
\end{equation}
Here we invoke the electric dipolar approximation $F_j(t)=\vec{\mu}_j\cdot\vec{E}(t)$ where $\vec{\mu}_j$ is the transition dipole moment and $\vec{E}(t)$ is the electric field of the incoming field. 
With the long wavelength approximation, we assume $F_j(t)=f(t)$ for all $j$ and choose $f(t)=A\sin(\omega_d t)\exp(-(t-t_d)^2/B^2)$ as a Gaussian light pulse.
Here $A$ is the pulse amplitude, $B$ is the duration of the pulse, and $t_d$ indicates the peak of the pulse.
In the frequency domain, the Fourier transform of $f(t)$ is a Gaussian distribution where $\omega_d$ is the central frequency and $1/\pi B$ is the spectral width.

\begin{figure}
\includegraphics{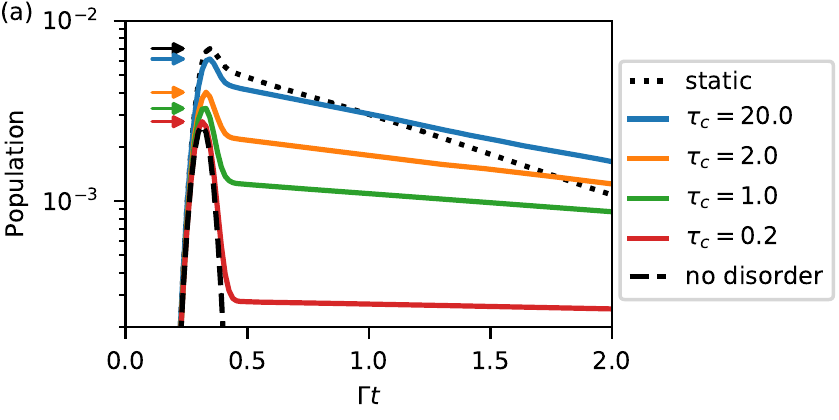}\\ 
\includegraphics{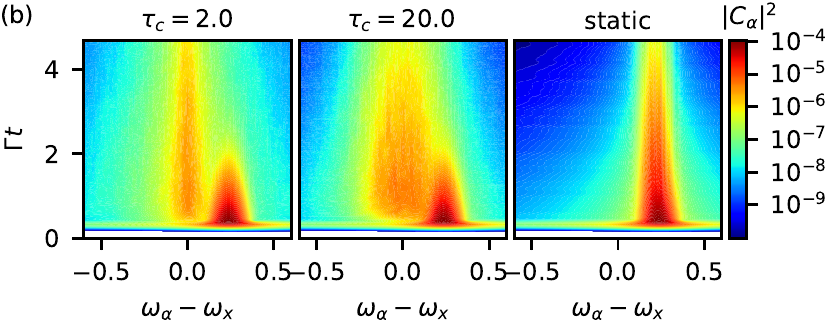}
\caption{Off-resonant pulsed excitation of a molecular ensemble experiencing disorder.
The initial state is the ground molecular state $\ket{G}$, the disorder amplitude is $\sigma=0.1$, and the driving frequency of the light pulse has a detuning $\omega_d-\omega_x=0.25$.
(a) Molecular excitation population as a function of time.
The maximal population $P_\mathrm{max}$ is denoted by arrows for different disorder cases.
After the pulsed excitation, the population dynamics show a biexponential decay. 
For the long-time dynamics, the static disorder case decays at the spontaneous emission rate $e^{-\Gamma t}$ (black dotted line), whereas the dynamic disorder cases lead to a subradiant decay rate ($\Gamma'<\Gamma$). 
The black dashed line is the case without disorder.
(b) Heat map of $|C_\alpha(t)|^2$ as a function of $\omega_\alpha-\omega_x$ and $t$ for $\tau_c=2,20$ and the static disorder case.
Elastic scattering is observed at the pulse frequency $\omega_\alpha-\omega_x=0.25$ and fluorescence is observed at the average molecule energy $\omega_\alpha-\omega_x=0$.
Note that the fluorescence emission corresponds to the long-time, subradiant population decay and that the fluorescence signal is narrower for $\tau_c=2$ versus $\tau_c=20$ (i.e. motional narrowing).
For the static disorder case, we observe only the elastic scattering emission.}\label{fig3}
\end{figure}

In what follows, we report results of calculation based on the model above using the same parameters as in Sec.~\ref{sec:superradiance}.
Before pumping, all molecules are initialized to be in the ground state $\ket{G}$ (i.e. $P(0)=0$).
The incoming pulse is weak ($A=5\times10^{-3}$) and the pulse frequency is off-resonant with a detuning $\omega_d-\omega_x=0.25$.
Moreover, we choose $t_d=100$ and the duration of the Gaussian pulse $B=25$ so that the spectral width in the frequency domain is smaller than the detuning ($1/\pi B<\omega_d-\omega_x$).
As such, in the absence of disorder, this off-resonant light pulse leads to a transient excitation population of the molecular ensemble, that disappears (together with the accompanying scattering signal) with the pulse at $\Gamma(t_d+B)\approx0.4$ (see the black dashed line in Fig.~\ref{fig3}(a)).
%(i.e. the emission spectrum has only a scattering peak at $\hbar\omega_\alpha-E_x=0.25$ and width $2/B=0.08$ (see black dotted line in Fig.~\ref{fig3}(b)).

%AN: It would be good to have a short paragraph here listing clearly what are the observables that you are about to show below, then refer to equations here when you discuss an observable.
%HTC: I thought I list them already. If this is about the Fig.3 caption, I have fixed that. 

\subsection{Including disorder enhances the maximal excitation population}
Fig.~\ref{fig3}(a) shows that, in the presence of disorder (both static and dynamic), the maximal value of the excitation population as induced by the off-resonant light pulse is enhanced (see $P_\text{max}$ as labeled by the arrows) relative to no disorder (black dashed lines).
For a fixed disorder amplitude $\sigma=0.1$, such an enhancement is the strongest for the case of static disorder (black dotted line), for which the maximal excitation population (black arrow) can be $3$ times larger than that in for the ordered system (black dashed line).
This observation can be rationalized by the fact that, in the presence of disorder, some molecules are closer to resonance with the incident radiation.
For dynamic disorder (solid lines), as $\tau_c$ decreases, the maximal value of the population becomes ever smaller and eventually approaches the no disorder result in the limit of very fast modulations (red line). 

\newparagraph{If we turn off the pulse fast enough at $t=t_d$, we can observe the superradiant decay followed by a subradiant decay at long times for dynamic disorder (see Fig.~\ref{fig8} in Appendix) and recover the same behavior as in Fig.~\ref{fig1} where the dynamics is started from a superradiant state.
This observation suggests that the molecular ensemble at $P_\text{max}$ is in the collective superradiant state.
Note that below we will focus on the off-resonant scattering of a Gaussian light pulse and, in this case, the superradiant decay is difficult to observe during the short time span that the pulse disappears.}

\subsection{Elastic scattering in the presence of static disorder}\label{sec:static_scattering} 
% Let us begin by focusing on the results for static disorder (black dashed line) in Fig.~\ref{fig3}.
The black dotted line in Fig.~\ref{fig3}(a) shows the time evolution of the excitation population following the pulse excitation of the molecular ensemble in the presence of static disorder. 
We notice that, following the incoming light pulse, the excitation population dynamics for static disorder decays at the single-molecule spontaneous emission rate $\Gamma$ at long times ($\Gamma t>0.5$).
This observation implies that, for static disorder, the light scattering process is dominated by a few (even one) molecules which are on resonance with the incoming light ($\tilde{\omega}_j\approx\omega_d$).
In other words, for static disorder, each of these molecules scatters the incoming pulse independently and there is no observation of collective coherence.

In Fig.~\ref{fig3}(b), we plot the energy distribution of the emitted light (i.e. the emission spectrum) as represented in our model by the population of the emitted photon states $|C_\alpha(t)|^2$. 
The right panel in Fig.~\ref{fig3}(b) shows this spectrum in the static disorder limit and the elastic scattering signal is observed in the frequency range centered at $\omega_\alpha\approx\omega_d$ (the driving frequency) at long times. 
Note that the lineshape here is averaged over $256$ realizations and, if we were to analyze one single realization, we would find a collection of much narrower streaks in the spectrum (each representing one elastic scattering event). 
In other words, the observed signal at $\tilde{\omega}_j\approx\omega_d$ represents an inhomogeneous average of many dynamic signals.

\subsection{Dynamic disorder: fluorescence emission at a subradiant rate}
Next, let us analyze the results for dynamic disorder (solid lines in Fig.~\ref{fig3}).
Following the off-resonant incident pulse, the population dynamics exhibits a biexpoential relaxation.
In the limit $\tau_c\rightarrow\infty$, the population dynamics can almost recover the elastic scattering in the presence of static disorder.
In the limit $\tau_c\rightarrow0$, the stochastic modulation becomes too fast for the molecules to interact with the incident pulse, so that the molecules cannot be efficiently excited leading to smaller maximal population (red line) as in the case without disorder. 
For the correlation time in the intermediate range $\tau_c\approx O(2\pi/\omega_d)$, the molecules can be excited, but cannot construct the molecular coherence, leading to a subradiant state.
As such, the long-time dynamics decays at a subradiant rate (which is slower than the spontaneous emission seen in the static disorder case).
This subradiant decay implies that, under dynamic disorder, the excitation energy is held for a longer time within the molecular subsystem and emission is slower.
% To rationalize this observation, we imagine that each molecule has an electronic transition frequency $\omega_j(t)$ that changes in time stochastically and can be excited by the incoming light when the molecule is on resonance $\omega_j(t)\approx\omega_d$. 
% If the stochastic modulation rate is faster than the molecular emission timescale, one expects to find that the molecules that have $\omega_j(t)\approx\omega_d$ at time $t$ are excited, and subsequently the excitation spectrum for those same molecules will be rapidly relax to a Gaussian distribution (centered at $\omega_x$) before any single-molecule emission occurs. 
% As such, changes in the excitation energies as induced by dynamic disorder can lead to interference among the molecular excited states and suppress collective emission.
% However, if the stochastic modulation becomes too fast for the molecules to interact with the incoming pulse, the molecules cannot be excited as much leading to smaller maximal population (red line). 

The corresponding emission spectrum is displayed in the \change{left} and middle panels of Fig.~\ref{fig3}(b).
% In Fig.~\ref{fig3}(b), these same phenomena are now analyzed from the point of view of the emitted radiation. 
Under dynamic disorder, the emission spectrum $|C_\alpha(t)|^2$ shows two components: the scattering component (S) centered at the external driving frequency ($\omega_\alpha-\omega_x=0.25$), and the fluorescence emission component (F) centered at the average molecular excitation energy ($\omega_\alpha-\omega_x=0$).
On the one hand, the scattering component decays quickly after the pulse excitation subsides, and its duration is independent of $\tau_c$ and remains almost the same as the case without disorder.
On the other hand, the fluorescence emission signal emerges mostly after the pulse and is clearly induced by dynamic disorder. 
The fluorescence emission component has a long lifetime ($\Gamma t>5$), which corresponds to the slow, subradiant decay of the excitation population.
Note that the linewidth of the fluorescence emission in the frequency domain becomes narrower as $\tau_c$ decreases, showing motional narrowing of the fluorescence emission component (as opposed to the elastic scattering at short times).

\begin{figure}
\includegraphics{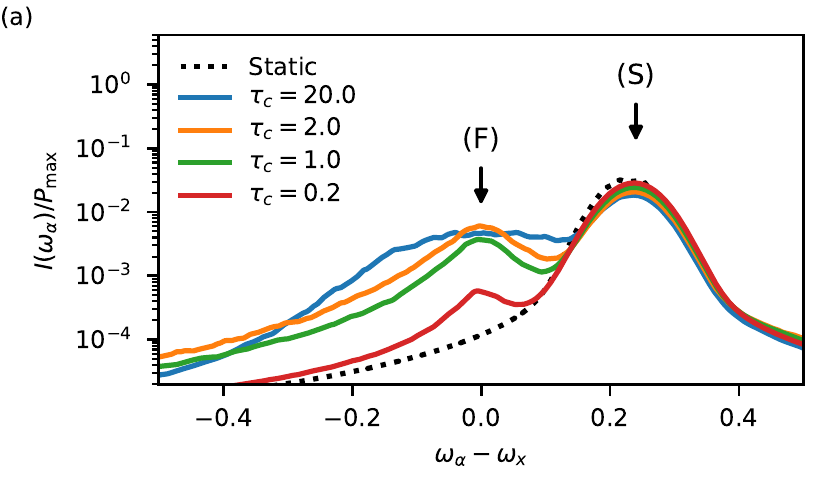} 
\includegraphics{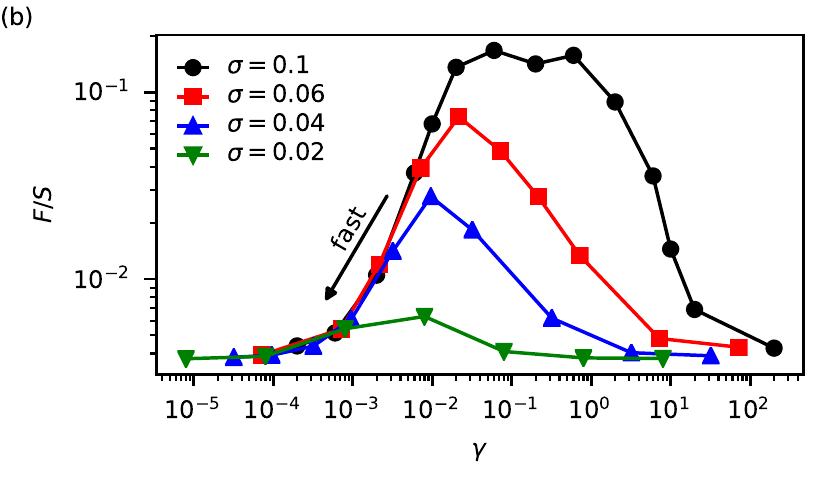} 
\caption{(a) The normalized cumulative emission spectrum $I(\omega_\alpha,t)/P_\mathrm{max}$ at $\Gamma t=2.5$ is plotted as a function of $\omega_\alpha-\omega_x$ for disordered molecular ensembles with the disorder amplitude $\sigma=0.1$.
The initial state is $\ket{G}$ and the pulse frequency is off-resonant $\omega_d-\omega_x=0.25$. 
For the dynamic disorder cases with the correlation times $\tau_c=20,2,1,0.2$ (solid lines), the emission spectrum shows the scattering peak (S) at the pulse frequency $\omega_\alpha=\omega_d$ and the fluorescence peak (F) at the average molecular transition frequency $\omega_\alpha=\omega_x$. 
The fluorescence peak becomes narrower when $\tau_c$ deceases, indicating motional narrowing.
In contrast, for static disorder (black dotted line), the emission spectrum shows only the scattering peak.
(b) The $F/S$ ratio as a function of the dephasing rate $\gamma=\sigma^2\tau_c$ is plotted for $\sigma=0.1, 0.06, 0.04, 0.02$. 
Note that dynamic disorder leads to an enhancement of the fluorescence emission in the range of intermediate modulation.
As expected, for long correlation times, when $\tau_c$ increases with a fixed $\sigma$, the $F/S$ ratio decreases.
Interestingly, for short correlation times ($\gamma<10^{-2}$), the $F/S$ ratio decreases again when $\gamma$ decreases.
Note that for meaningful fluorescence signals, if the pulse is off-resonant, one requires some dynamic disorder ($\tau_c<\infty$) to allow the molecular frequency to align with the pulse frequency; however, at the same time, emitting a photon requires a finite amount of time and the fluorescence signal decreases when $\tau_c$ becomes very small. 
 }\label{fig4}
\end{figure}

\subsection{Fluorescence/scattering ratio turnover in the intermediate modulation regime}\label{sec:turnover}
The results discussed above suggest that the fluorescence, unlike the scattering component, is affected by the dynamics of the disorder.
To better quantify the relative importance of these molecular response components, we show in Fig.~\ref{fig4}(a) the cumulative emission spectra (Eq.~\eqref{eq:Icum}) at the end of the simulation time i.e. $\Gamma t=5$.
Here we normalize the cumulative emission by the maximal value of the molecular population ($P_\mathrm{max}$) and denote the yield at $\omega_\alpha-\omega_x=0.25$ as the scattering peak (S)  and the yield at  $\omega_\alpha-\omega_x=0$ as the fluorescence peak (F).
We find that the scattering components of the normalized cumulative emission remain almost the same for different values of $\tau_c$, confirming that the yield of the elastic scattering is not sensitive to disorder in the molecular system.
In contrast, the fluorescence component emerges in the presence of dynamic disorder: a wide Gaussian distribution for slow modulation ($\tau_c=20$) and a narrow Lorentzian distribution for fast modulation ($\tau_c=0.2$) due to motional narrowing.  
Interestingly, in both the fast modulation limit ($\tau_c\rightarrow 0$) and the static disorder limit ($\tau_c\rightarrow\infty$), the fluorescence peak disappears. 

% Turnover
In order to quantitatively compare the contribution of the scattering and fluorescence components, we fit the cumulative emission spectrum $I(\omega_\alpha)$ to a bimodal distribution.
In practice, we first fit the scattering peak to a Gaussian distribution (i.e. $I_\text{sct}(\omega_\alpha)\approx a'e^{-(\omega_\alpha-\omega_d)^2/b'^2}$), and then second the rest of the emission is considered fluorescence  ($I_\text{flu}(\omega_\alpha)=I(\omega_\alpha)-I_\text{sct}(\omega_\alpha)$).
%HTC: (Note that we cannot simply fit the fluorescence peak to a Lorentzian distribution because the fluorescence component turns into a Gaussian distribution in the slow modulation limit.)
With these fitted components, we calculate the total contribution of the scattering and fluorescence components by $F\equiv\int{d\omega_\alpha}I_\text{flu}(\omega_\alpha)$ and $S\equiv\int{d\omega_\alpha}I_\text{sct}(\omega_\alpha)$ respectively.

Fig.~\ref{fig4}(b) shows the ratio $F/S$ as a function of the dephasing rate $\gamma=\sigma^2\tau_c$ for different disorder amplitudes $\sigma$. 
Let us first consider the case $\sigma = 0.1$.
We find a maximum (or really a plateau) in the $F/S$ ratio over the
range $\gamma \in [10^{-2}, 10^{0}]$. 
Otherwise, $F/S$ decays as $\gamma \rightarrow \infty$ ($\tau_c \rightarrow \infty$) and $\gamma \rightarrow 0$ ($\tau_c \rightarrow 0$).
%We observe a turnover of the  $F/S$ ratio when decreasing $\tau_c$ from the static disorder limit ($\tau_c\rightarrow\infty$).
%For long correlation times ($\gamma>10^{-2}$), as $\tau_c$ decreases, the  $F/S$ ratio increases (and plateaus around $F/S=0.15$ for $\sigma=0.1$).
%For short correlation times ($\gamma<10^{-2}$) , as $\tau_c$ decreases,  the  $F/S$ ratio decreases and converge for different values of $\sigma$. 
These same conclusions are qualitatively found for different $\sigma$ values as well. 
Such a turnover behavior suggests that observing the fluorescence signal as induced by an off-resonant incoming pulse requires the dynamic disorder parameters ($\sigma$ and $\tau_c$) to be in an intermediate regime.
Namely, the stochastic process must be fast enough to modulate the molecular excitation before emitting an photon; however, at the same time, the stochastic process cannot be too fast for the molecules to absorb the incoming photon. 
From the perspective of energy conservation, the fluorescence response is essentially an inelastic scattering process with the excess energy dissipated to the environmental fluctuations. This relaxation channel is maximized when these fluctuations are dominated by timescales that match the frequency difference $\omega_d-\omega_x$.

% For $\tau_c=20$ (blue), the relatively larger fluorescence component after the pulse is corresponding to the faster long-time decay rate relative to other smaller $\tau_c$ cases.

% That being said, we notice that, as $\tau_c$ decreases, the fluorescence component can be suppressed by orders of magnitude.
% Such suppression can be attributed to motional narrowing in the sense that the effective width of the molecule excitation becomes smaller so that the external field cannot excite the molecules as much. 
% Note that, for $\tau_c=0.2$, the scattering peak is almost identical with the emission spectrum without disorder and the fluorescence peak is really small--- similar to the case without disorder.
% Therefore, we conclude that dynamic disorder can induce the long-time fluorescence emission given that the correlation time is in the range $2\pi\hbar/E_x<\tau_c<1/\Gamma$.

\subsection{Fast modulation leads to large participation ratio}
Next consider the collective aspect of the observed molecular response and the dependence of the emitted radiation on the molecular number $N$.
In order to estimate how many quantum emitters are excited in a molecular ensemble, we can calculate the normalized participation ratio of the wavefunction of the molecular subsystem,
\begin{equation}\label{eq:PR}
\mathrm{PR}
=\left<\frac{\left[\sum_{j}|\left<X_j|\psi(t)\right>|^2\right]^2}{\sum_{j}|\left<X_j|\psi(t)\right>|^4}\right>
% =\left<\frac{\left[\sum_{j}|C_j|^2\right]^2}{\sum_{j}|C_j|^4}\right>
=\left<\frac{P(t)^2}{\sum_{j}|C_j(t)|^4}\right>.
\end{equation}
Note that, since the wavefunction of the molecular subsystem is not necessary normalized (i.e. $P(t)=\sum_{j}|C_j(t)|^2\neq1$ for the pulsed excitation dynamics), Eq.~\eqref{eq:PR} is defined as if we first normalize the subsystem wavefunction $\tilde{C}_j(t)=C_j(t)/\sqrt{P(t)}$ and then calculate the participation ratio using the standard definition\cite{biella_subradiant_2013} $\mathrm{PR}=\left<\frac{1}{\sum_{j}|\tilde{C}_j(t)|^4}\right>$.
For completely delocalized states $\tilde{C}_j=\frac{1}{\sqrt{N}}$ for all $j$, we have $\mathrm{PR}=N$, which indicates that the wavefunction is delocalized throughout $N$ molecules.
For a completely localized state, $\mathrm{PR}=1$.

\begin{figure}
\includegraphics{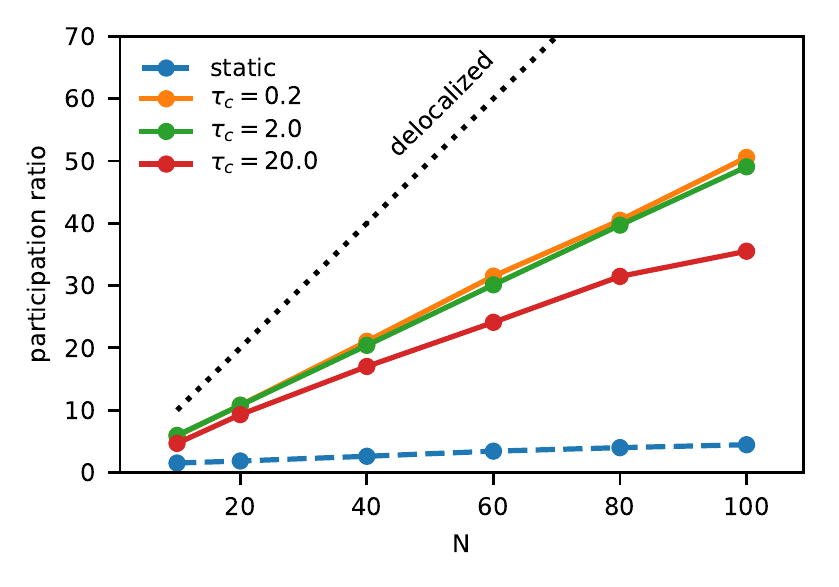}
\caption{The participation ratio at $\Gamma t= 2.5$ as a function of the number of emitters $N$ for $\tau_c=0.2,2,20$. 
We choose the disorder amplitude to be $\sigma=0.1$.
Here $PR$ increases as $\tau_c$ decreases and reaches the maximum $PR=N/2$ in the fast modulation limit ($\tau_c=0.2$). 
Note that $PR=N/2$ implies that the wavefunction involves half of the single excitation states. 
}\label{fig_IPR}
\end{figure}

Fig.~\ref{fig_IPR} shows the normalized participation ratio of the molecular subsystem wavefunction as a function of $N$. 
Here we focus the long-time wavefunction ($\Gamma t=5$) when the elastic scattering signal vanishes and the fluorescence emission remains.
For static disorder, as expected, $PR\rightarrow1$  and the molecular excitation is formed by only one (or few) single excitation state. 
For short correlation times ($\tau_c=2, 0.2$), we find $PR\approx N/2$ and scales linearly with $N$, implying that nearly half of the molecules are involved, i.e. the wavefunction is a combination of $N/2$ single excitation states $\ket{X_j}$.
We note that, as $\tau_c$ becomes larger ($\tau_c=20$), $PR$ decreases and the wavefunction is composed of fewer single excitation states.
This result clearly implies that including dynamic disorder enhances the collectivity of the molecular excitation as induced by an off-resonant incoming pulse. This observation is consistent with the result of Fig.~\ref{fig1}, where we found that faster dynamic disorder more efficiently preserve superradiance response).

\subsection{N-dependence of the emission spectrum}
\begin{figure}
\includegraphics{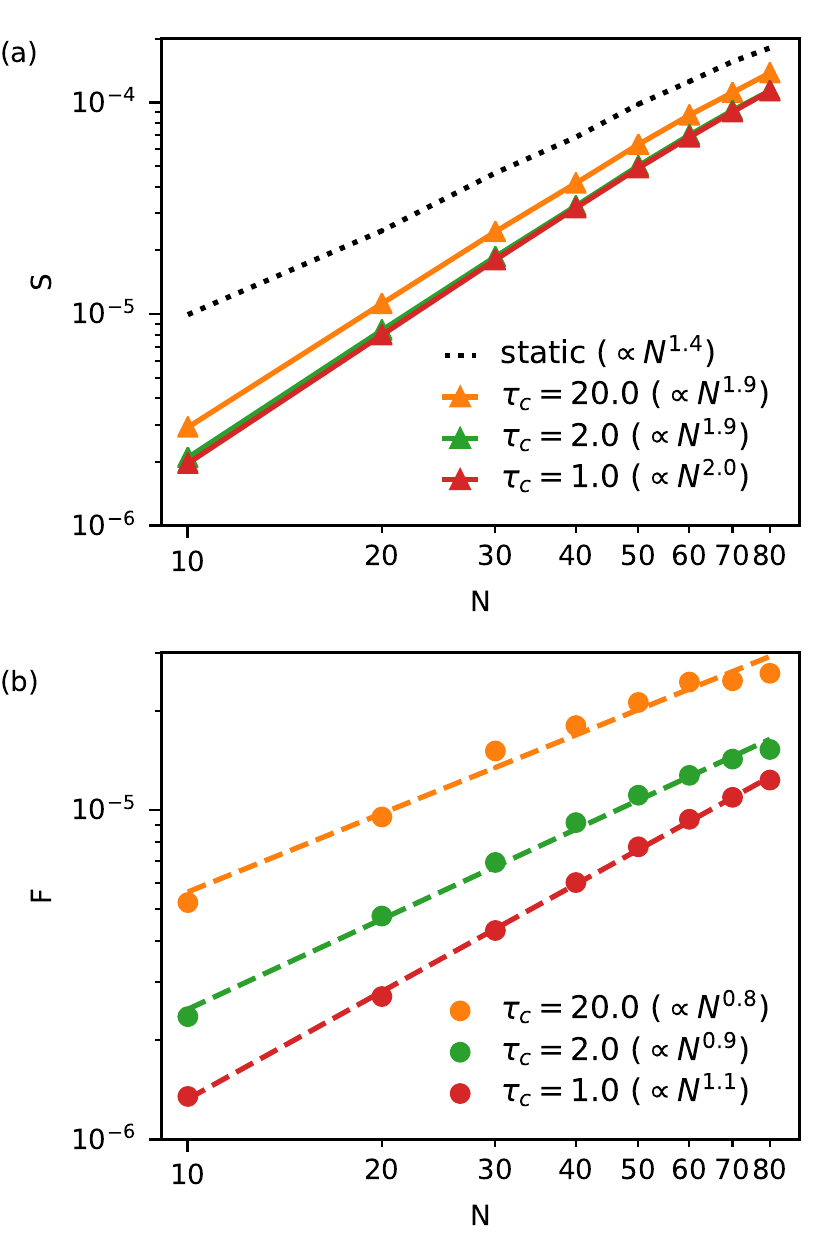}
\caption{\change{The total contribution of the elastic scattering (a) and fluorescence emission (b) as a function of the number of molecules ($N$) in a log-log scale. 
The initial state is $\ket{G}$ and the frequency of the light pulse is off-resonant $\omega_d-\omega_x=0.25$.
The disorder amplitude is $\sigma=0.1$ and we vary the correlation time $\tau_c=20,2,1$. 
For dynamic disorder, the scattering intensity scales quadratically with $N$, which is a signature of the collective superradiant emission, and the fluorescence intensity scales linearly with $N$.
In contrast, for static disorder (black dashed line), the scattering intensity scales as $N^{1.4}$, implying that the scattering signal has contributions both from single-molecule and collective emission.
}
}\label{fig6}
\end{figure}

Finally, we consider the $N$-dependence of the $S$ and $F$ contributions in the emission spectrum.
The data is plotted in Fig.~\ref{fig6}.
Here we choose dynamic disorder with $\sigma=0.1$ and $\tau_c=0.2, 1, 2$ to be in the parameter range where the fluorescence signal can be clearly observed.
We find that the elastic scattering signals $S$ has a quadratic dependence on $N$ ($S\propto N^2$ in Fig.~\ref{fig6}(a)) and the fluorescence emission signals $F$ scales linearly with $N$ ($F\propto N$ in Fig.~\ref{fig6}(b)). 

To understand $N$-dependence of the signal, we follow the Kramers–Heisenberg–Dirac (KHD) formalism\cite{tannor_introduction_2006} and express the ratio between the incoming and emission intensities in terms of the scattering cross section
\begin{equation*}
\frac{I(\omega_\alpha)\omega_{d}}{I_{in}\omega_{\alpha}}=\sigma_{f\leftarrow i}\left(\omega_{d},\omega_{\alpha}\right)
\end{equation*}
Here the incident light has the frequency $\omega_d$ and the intensity $I_{in}$.
Next, we evoke the second-order perturbation approach as in Ref.~\citenum{tannor_introduction_2006} and the scattering cross section can be written in the sum-over-states expression\cite{heller_semiclassical_1981}:
\begin{equation}\label{eq:sum-over-states}
    \sigma_{f\leftarrow i}\left(\omega_d,\omega_\alpha\right)=\frac{\omega_d\omega_\alpha}{4\pi c^{2}}\left|\sum_{k}\frac{\left\langle \phi_{f}\right|\hat{V}\left|\phi_{k}\right\rangle \left\langle \phi_{k}\right|\hat{V}\left|\phi_{i}\right\rangle }{\omega_{i}+\omega_{d}-\omega_{k}+i\gamma_{k}}\right|^{2}.
\end{equation}
Here $\ket{\phi_{i}}$ and $\ket{\phi_{f}}$ are the initial and final electronic states respectively and, for our purpose, we choose $\ket{\phi_{i}}=\ket{\phi_{f}}=\ket{G}$ \change{and $\hbar\omega_G$ is the total ground state energy}.
Eq.\eqref{eq:sum-over-states} sums over all the intermediate state $\ket{\phi_{k}}$ (with the frequency $\omega_k$ and the lifetime $\gamma_{k}$) that are involved in the light scattering process from $i$ to $f$.
In the following, we consider the elastic scattering and fluorescence emission signals in this formalism that result from different intermediate states.

\newparagraph{
\emph{(i) Elastic scattering, fast modulation limit:} We first consider the fast modulation limit in which the molecular excitation energy can fluctuate rapidly and cover almost the entire disorder spectrum. 
Thus, at any instant, each single molecule should have a fraction of the probability distribution ($0<\xi<1$) to be resonant with the incoming light ($\omega_j(t)\approx\omega_d$).
Note that the parameter $\xi$ should depend on the disorder of the molecular ensemble and the detuning of the incoming pulse--but not $N$.
For $N$ identical molecules under the same pulse excitation, such a rapid fluctuation builds up molecular coherence and leads to the collective superradiant state, i.e. $\ket{\phi_{k}}=\sqrt{\frac{\xi}{N}}\sum_{j}^{N}\ket{X_{j}}$ (with the total excitation probability $\xi$).
With this intermediate state, the scattering signal intensity can be estimated by
\begin{equation}\label{eq:intensity_2}
I(\omega_d)\propto\left|\sum_{j'}^{N}\sum_{j''}^{N}\frac{\xi}{N}\frac{\bra{G}\hat{V}\ket{X_{j'}}\bra{X_{j''}}\hat{V}\ket{G}}{\omega_G+i\Gamma}\right|^2\propto \xi^2N^2
\end{equation}
This finding explains the quadratic $N$-dependence of the scattering signals (see Fig.~\ref{fig6}(a)). 
}

\change{
\emph{(ii) Elastic scattering, static disorder limit:} Next, we focus on the scattering intensity in the static disorder limit and notice that all the observed emission signals are centered at $\omega_d$ (as shown in Fig.~\ref{fig4}). As we discussed in Sec.\ref{sec:static_scattering}, one can imagine a fraction of molecules ($\xi N$ for $0<\xi<1$) are on resonance with the incoming light ($\tilde{\omega}_j\approx\omega_d$).
On the one hand, the excitation pulse can build coherence among these molecules and form the superradiant state, implying that the signal intensity scales as $N^2$.
On the other hand, there is still some static disorder among the molecular excitation energies, which will inevitably lead to a loss of coherence such that the molecules will emit individually, and the therefore the signal will be proportional to $N$.
The competition between these mechanisms explains the intermediate $N$-dependence between linear and quadratic scaling of the scattering signal in the case of static disorder ($\propto N^{1.4}$ as shown in Fig.~\ref{fig6}(a)).}

\change{\emph{(iii) Fluorescence emission:}} As we discussed in Sec.~\ref{sec:turnover}, the fluorescence emission is essentially inelastic scattering through a subradiant intermediate state.
% To describe the subradiant states, we now employ the effective non-Hermitian Hamiltonian $\hat{H}_\mathrm{eff}=\hat{H}_M-i\frac{\Gamma}{2}\hat{Q}$ where $\bra{X_i}\hat{Q}\ket{X_j}=1$. 
At the same time, Fig.~\ref{fig_IPR} suggest that the participation ratio is $N/2$ in fast modulation limit, i.e. half of the molecules are involved in the scattering process and have the average excitation population $\frac{2}{N}$. 
In this fast modulation limit, we assume the subradiant wavefunction takes the form
$|{\tilde{\psi}}\rangle=\sum_{j'=1}^{N/2}\sqrt{\frac{2}{N}}e^{i\varphi_{j'}}\ket{X_{j'}}$ where $\varphi_{j'}$ is an arbitrary phase.
The subradiant state has energy around $\omega_x$ and inverse lifetime $\Gamma'<\Gamma$.
Therefore, the scattering intensity through the subradiant intermediate state can be estimated by
\begin{align}
I(\omega_x)
&\propto\left<\left|\sum_{j',j''}^{ N/2}\frac{2}{N}\frac{\bra{G}\hat{V}\ket{X_{j'}}\bra{X_{j''}}\hat{V}\ket{G}}{\omega_G+\omega_d-\omega_x+i\Gamma'}e^{i(\phi_{j'}-\phi_{j''})}\right|^2\right> \label{eq:F_signal}\\
&\propto\sum_{j'=1}^{ N/2}\left|\frac{\bra{G}\hat{V}\ket{X_{j'}}\bra{X_{j'}}\hat{V}\ket{G}}{\omega_G+\omega_d-\omega_x+i\Gamma'}\right|^2
\propto N.
\end{align}
Note that we expand the squared norm in Eq.~\eqref{eq:F_signal} and, on average, the cross terms with an arbitrary phase difference $e^{i(\varphi_{j'}-\varphi_{k'})}$ should cancel out, which is the key for the fluorescence emission to have the linear scaling with $N$, rather than $N^2$ dependence.

\section{Conclusion}\label{sec:conclusion}
In this work, we have investigated the collective response of a molecular ensemble of quantum emitters exposed to environmental dynamic disorder with various correlation time scales.
Our results show that, in the short correlation time limit, dynamic disorder can effectively recover the coherent response of the molecular ensemble leading to fast relaxation at the superradiant rate; this coherence is suppressed in the static disorder limit.
More interestingly, recovery of the superradiant decay in the excitation population dynamics is concomitant with motional narrowing of the emission spectrum.

Following an off-resonant incident pulse, if dynamic disorder has an appropriate correlation timescale that allows for energy exchange between the incident pulse and the environmental fluctuations, the molecular ensemble can relax at a slow, subradiant rate, leading eventually to inelastic fluorescence emission component at long times.
As a result, the subradiant state of the molecular ensemble is a collective excitation state (i.e. involve many single excitations) that can live for a long time due to dynamic disorder.
We also show that, the fluorescence component scales linearly with the number of the quantum emitters, suggesting a distinct (incoherent) collective feature of the subradiant state (compared to the quadratic scaling of the elastic scattering component).

These results suggest that accounting for environmental disorder effects in term of stochastic modulation of the electronic transition frequency is important for collective excitation and emission. 
That being said, there are many assumption in our collective excitation model that can be scrutinized.
First, we assume a symmetric Gaussian stochastic modulation that has an equal probability for increasing and decreasing the excitation energy (effectively infinite temperature environment). 
At low temperature $kT<\sigma$, the stochastic random variable should show the consequence of detailed balance and recover the correct thermal equilibrium.\footnote{\change{In a quantum description of the thermal environment with a generic bath operator $\hat{B}(t)$, such correlation functions satisfy the property $\tilde{C}_{BB}(\omega) = e^{\beta\hbar\omega}\tilde{C}_{BB}(-\omega)$ where $\tilde{C}_{BB}(\omega)=\int_{-\infty}^{\infty}{dt}e^{i\omega t}\langle\hat{B}(t)\hat{B}(0)\rangle$ is the Fourier transform of the correlation function and the factor $e^{\beta\hbar\omega}$ accounts for detailed balance. Note that the classical stochastic modeling used in our model does not include this factor (i.e. a high temperature approximation, $e^{\beta\hbar\omega}\rightarrow1$).}}
Second, the coupling to the radiation field continuum is assumed to be identical for all the molecules, which ignores spatial dependence and orientation disorder.
Third, we conveniently neglect the influence of  molecular vibrations and strong coupling between the vibrational modes and photon states, which can be taken into account (at least heuristically) in the framework of macroscopic quantum electrodynamics.\cite{wang_theory_2020,lee_theory_2021,wang_coherent--incoherent_2020}
Finally, we make the wide band approximation for the radiative relaxation channels, which is valid only when the edges of the continuum are far from the molecular excitation energy.
More generally, one should be able to employ a semiclassical model (for example the Maxwell-Bloch equation) for a more realistic model system.
Future research into these generalizations is currently underway.

Looking forward, restoring the molecular coherence and constructing a collective behavior using dynamic disorder would be useful for many applications in the field of nanophotonics. 
For example, concerning many recent interests in cavity polaritons in physical chemistry community\cite{mandal_polariton-mediated_2020,engelhardt_unusual_2022,herrera_disordered_2022,smith_exact_2021}, one often probes the responses of the molecules within an optical cavity through the upper and lower polariton states under the influence of the environmental disorder. 
Can we manipulate the lifetime of the polariton states by changing the timescale of the environmental fluctuations? 
Can we use dynamic disorder as a tuning knob for controlling chemical reactions within an optical cavity?
These directions of investigation will be taken up in a future work.

\section*{Acknowledgements}
\change{
This work has been supported by the U.S. Department of Energy, Office of Science, Office of Basic Energy Sciences, under Award No. DE-SC0019397 (JES) and the U.S. National Science Foundation under Grant No. CHE1953701 (A.N.), and the Air Force Office of Scientific Research under Grant No. FA9550-22-1-0175 (M.S.). It also used resources of the National Energy Research Scientific Computing Center (NERSC), a U.S. Department of Energy Office of Science User Facility operated under Contract No. DE-AC02-05CH11231.}

\appendix
\section{Generating Gaussian stochastic variables}\label{sec:generate}
In this paper we have implemented a Gaussian random process $x(t)$ with zero mean ($\left<x(t)\right>=0$) and the exponential correlation function $\left<x(t_1)x(t_2)\right>=\sigma^{2}e^{-|t_1-t_2|/\tau_{c}}$. Such a process was simulated following Ref.~\citenum{gillespie_exact_1996,rybicki_notes_1994}.
For an ordered set of discrete times $\{t_i\}$ ($t_1<t_2<\cdots<t_n$), we let $x_i=x(t_i)$ be the values of the Gaussian random process.
The joint probability distribution of $\{x_1,\cdots,x_n\}$ can be expressed as a product of conditional probability
\begin{equation}
    \mathrm{Prob}(x_1,\cdots,x_n)=\mathrm{Prob}(x_1)\prod_{i=2}^{n}\mathrm{Prob}(x_i|x_{i-1}).
\end{equation}
Here the initial probability distribution of $x_1$ is 
\begin{equation}
    \mathrm{Prob}(x_1)=\frac{1}{\sqrt{2\pi\sigma^2}}\exp\left[-\frac{x_1^2}{2\sigma^2}\right]\label{eq:initial}
\end{equation}
and the conditional probability distribution of $x_i$ given the value $x_{i-1}$ is 
\begin{equation}
    \mathrm{Prob}(x_i|x_{i-1})=\frac{1}{\sqrt{2\pi\sigma^2(1-r_{i-1}^2)}}\exp\left[-\frac{(x_i-r_{i-1}x_{i-1})^2}{2\sigma^2(1-r_{i-1}^2)}\right]\label{eq:conditional}
\end{equation}
where $r_i=e^{-(t_{i+1}-t_{i})/\tau_c}$ for $1<i<n-1$.
If we let $t_{i}=idt$, $r_i=e^{-dt/\tau_c}$ does not depends on $i$.
Therefore, the conditional probability distribution of $x_i$ is a Gaussian distribution with mean $\bar{x}_i=x_{i-1}e^{-dt/\tau_c}$ and variance $\sigma^2(1-e^{-2dt/\tau_c})$.
We notice that, in the slow modulation limit (large $\tau_c$), the mean $x_{i-1}e^{-dt/\tau_c}\rightarrow x_{i-1}$ and the variance $\sigma^2(1-e^{-2dt/\tau_c})\approx\sigma^22dt/\tau_c\rightarrow0$, so that $x(t)$ becomes time-independent (static).

With this Markov property, we can generate Gaussian stochastic variables $\Omega_j(t)$ for each molecule as follows:
\begin{enumerate}
    \item Choose the initial value $\Omega_j(t=0)$ (i.e. $i=0$) from the Gaussian distribution in Eq.~\eqref{eq:initial},
    \item Calculate the mean $\bar{\Omega}_j=\Omega_j(t_{i-1})e^{-dt/\tau_c}$ for the next $i$,
    \item Choose $\Omega_j(t_{i})$ from a Gaussian distribution with the mean $\bar{\Omega}_j$ and the variance $\sigma^2(1-e^{-2dt/\tau_c})$,
    \item Go back to Step 2 for the next $i$.
\end{enumerate}

\newparagraph{
% time-dependent perturbation theory
\section{Time-dependent perturbation theory with dynamic disorder}\label{sec:TDPT}
In this section, we derive the first order approximation of the excitation population using time-dependent perturbation theory.\cite{fetter_quantum_2003,nitzan_chemical_2006} 
We let $\hat{H}_{\text{eff}}=\hat{H}_{0}+\hat{V}(t)$ where  $V(t)=\sum_{j}\Omega_{j}(t)|X_{j}\rangle\langle X_{j}|$ and $\hat{H}_{0}=\sum_{j}\omega_x\left|X_{j}\right\rangle \left\langle X_{j}\right|-i\frac{\Gamma}{2}\sum_{jk}\left|X_{j}\right\rangle \left\langle X_{k}\right|$. 
The electronic state wavefunction $|\phi(t)\rangle=\sum_jC_j(t)|X_j\rangle$ can be propagated by $|\phi(t)\rangle=\hat{U}(t)|\phi(0)\rangle$ where $|\phi(0)\rangle$ is the initial state at $t=0$.
Here, the propagator in the Schrodinger picture can be expanded in terms of $\hat{V}_I(t)=e^{i\hat{H}_{0}t}\hat{V}(t)e^{-i\hat{H}_{0}t}$ 
\begin{equation}\label{eq:propagator_all}
\hat{U}(t)=e^{-i\hat{H}_{0}t}-ie^{-i\hat{H}_{0}t}\int_{0}^{t}dt_{1}\hat{V}_{I}(t_{1})-e^{-i\hat{H}_{0}t}\int_{0}^{t}dt_{1}\int_{0}^{t_{1}}dt_{2}\hat{V}_{I}(t_{1})\hat{V}_{I}(t_{2})+\cdots
\end{equation}
and we can write $\hat{U}(t)=\sum_{n=0}^{\infty}\hat{U}^{(n)}(t)$ where $n$ indicates the number of $\hat{V}_I$ operators.

The unperturbed propagator ($n=0$) can be expressed as
\begin{equation}\label{eq:propagator_0}
    \hat{U}^{(0)}(t)=e^{-i\hat{H}_{0}t}=e^{-i\omega_{x}t}\left(\sum_{k=1}^{N-1}\hat{D}_{k}+e^{-\frac{N\Gamma}{2}t}\hat{B}\right)
\end{equation}
where $\hat{B}=\left|S\right\rangle \left\langle S\right|$ and $\hat{D}_{k}=\left|d_{k}\right\rangle \left\langle d_{k}\right|$. 
Here $\left|S\right\rangle$ and $\left|d_{k}\right\rangle$ are the eigenstates of $\hat{H}_{0}$:  $\left|S\right\rangle =\frac{1}{\sqrt{N}}\sum_{j=1}^{N}\left|X_{j}\right\rangle $ is the fully symmetric state which corresponds to a complex-valued eigenvalue $\omega_{x}-i\frac{N\Gamma}{2}$; $\{\left|d_{k}\right\rangle|k=1,\cdots,N-1\}$ are $N-1$ degenerate eigenstates that have a real-value eigenvalue $\omega_{x}$, i.e. the dark states of $\hat{H}_0$.
Within the degenerate dark state subspace, we choose all dark states to be orthonormal to each other $\left\langle d_{k}|d_{k'}\right\rangle =\delta_{kk'}$ and orthogonal to the superradiant state $\left\langle d_{k}|S\right\rangle =0$.
Next, we plug the unperturbed propagator Eq.~\ref{eq:propagator_0} into the first order propagator in Eq.~\ref{eq:propagator_all}
\begin{align}\label{eq:propagator_1}
\hat{U}^{(1)}(t)
=-ie^{-i\omega_{x}t}\int_{0}^{t}dt_{1}	&\left(\sum_{k=1}^{N-1}\sum_{k'=1}^{N-1}\hat{D}_{k}\hat{V}(t_{1})\hat{D}_{k'}+e^{-\frac{N\Gamma}{2}(t-t_{1})}\sum_{k'=1}^{N-1}\hat{B}\hat{V}(t_{1})\hat{D}_{k'}\right. \nonumber\\
&\left.+e^{-\frac{N\Gamma}{2}t_{1}}\sum_{k=1}^{N-1}\hat{D}_{k}\hat{V}(t_{1})\hat{B}+e^{-\frac{N\Gamma}{2}t}\hat{B}\hat{V}(t_{1})\hat{B}\right)
\end{align}
With this approximate propagator, the time evolution of the electronic state can be calculated by $\left|\phi(t)\right\rangle \approx (\hat{U}^{(0)}(t)+\hat{U}^{(1)}(t))\left|\phi(0)\right\rangle$.
As we assume the initial state to be $\left|\phi(0)\right\rangle=\left|S\right\rangle$, the first two terms in Eq.~\eqref{eq:propagator_1} are zero ($\hat{D}_k|S\rangle=0$) and the electronic state can be written as 
\begin{equation}
    \left|\phi(t)\right\rangle\approx C_{s}^{(0)}(t)\left|S\right\rangle. +C_{s}^{(1)}(t)\left|S\right\rangle 
    +\sum_{k=1}^{N-1}C_{k}^{(1)}(t)\left|d_{k}\right\rangle 
\end{equation}
Here the zeroth order coefficient is
\begin{equation}
    C_{s}^{(0)}(t)=e^{-i\omega_{x}t}e^{-\frac{N\Gamma}{2}t}
\end{equation}
and the first order coefficients are given by
\begin{align}
    C_{s}^{(1)}(t)&=-ie^{-i\omega_{x}t}e^{-\frac{N\Gamma}{2}t}\frac{1}{N}\int_{0}^{t}dt_{1}\sum_{j}\Omega_{j}(t_{1}) \\
    C_{k}^{(1)}(t)&=-ie^{-i\omega_{x}t}\frac{1}{\sqrt{N}}\int_{0}^{t}dt_{1}e^{-\frac{N\Gamma}{2}t_{1}}\sum_{j}d_{k}^{j}\Omega_{j}(t_{1})
\end{align}
Finally, we can take the ensemble average of the molecular excitation population $P(t)=\langle|\phi(t)|^2\rangle$ and find:
\begin{equation}\label{eq:Pall}
    P(t)=\left\langle|C_{s}^{(0)}(t)|^{2}\right\rangle+2\left\langle\text{Re}\left(C_{s}^{(0)}(t)^{\dagger}C_{s}^{(1)}(t)\right)\right\rangle+\left\langle|C_{s}^{(1)}(t)|^{2}\right\rangle+\sum_{k=1}^{N-1}\left\langle|C_{k}^{(1)}(t)|^{2}\right\rangle
\end{equation}
All the contributions are evaluated explicitly as follows:
\begin{enumerate}
    \item $\langle|C_{s}^{(0)}(t)|^{2}\rangle$ yields the superradiant decay of the molecular ensemble without disorder. We define the zeroth order term as 
    \begin{equation}\label{eq:Ps0}
        P_s^{(0)}(t)\equiv\left\langle|C_{s}^{(0)}(t)|^{2}\right\rangle =e^{-N\Gamma t}
    \end{equation}
    \item The cross term $C_{s}^{(0)}(t)^{\dagger}C_{s}^{(1)}(t)$ is purely imaginary, i.e. $\text{Re}\left(C_{s}^{(0)}(t)^{\dagger}C_{s}^{(1)}(t)\right)=0$.
    \item $\langle|C_{s}^{(1)}(t)|^{2}\rangle$ leads to an integral of the two-time correlation function of Gaussian stochastic random variable
    \begin{align}
        \left\langle \left|C_{s}^{(1)}(t)\right|^{2}\right\rangle &=\frac{e^{-N\Gamma t}}{N^{2}}\sum_{j}\int_{0}^{t}dt_{1}^{\prime}\int_{0}^{t}dt_{1}\left\langle \Omega_{j}(t_{1}^{\prime})\Omega_{j}(t_{1})\right\rangle
    \end{align}
    Here we can carry out the integration analytically and define the contribution as (let $\gamma=\sigma^2\tau_c$) 
    \begin{equation}\label{eq:Ps1}
          P_s^{(1)}(t)\equiv\frac{2}{N}e^{-N\Gamma t}\gamma\left(t+\tau_{c}e^{-t/\tau_{c}}-\tau_{c}\right).
    \end{equation}
    We note that $P_s^{(1)}(0)=0$ and $P_s^{(1)}(t\rightarrow\infty)=0$, implying that this term contributes only to the transient dynamics and does not affect the short-time and long-time behaviors. We also find that the maximal value of $P_s^{(1)}(t)$ around $t\approx\frac{1}{N\Gamma}$ so that the contribution of this term $P_s^{(1)}(t=\frac{1}{N\Gamma})=\frac{2}{N^2\Gamma e}$ is relatively small when $N$ is large.
    \item To evaluate $\sum_{k=1}^{N-1}\left\langle|C_{k}^{(1)}(t)|^{2}\right\rangle$, we first notice that, since we choose the dark states to be orthonormal (i.e. $\sum_{j}d_{k}^{j*}d_{k}^{j}=1$ for all $k=1,\cdots, N-1$), $\left\langle|C_{k}^{(1)}(t)|^{2}\right\rangle$ does not depend on $k$ 
    \begin{equation}
        \left\langle |C_{k}^{(1)}(t)|^{2}\right\rangle =\frac{\sigma^2}{N}\int_{0}^{t}dt_{1}^{\prime}\int_{0}^{t}dt_{1}e^{-\frac{N\Gamma}{2}(t_{1}+t_{1}^{\prime})}e^{-|t_{1}-t_{1}^{\prime}|/\tau_{c}}
    \end{equation}
    This integration can be carried out using integration by parts 
    \begin{equation}
        \left\langle |C_{k}^{(1)}(t)|^{2}\right\rangle = \frac{2\sigma^{2}}{N^{2}\Gamma}\left[\frac{1}{q}e^{-N\Gamma t}+\frac{1}{r}-\frac{N\Gamma}{rq}e^{-rt}\right]
    \end{equation}
    where $r=\frac{N\Gamma}{2}+\frac{1}{\tau_{c}}$ and $q=\frac{N\Gamma}{2}-\frac{1}{\tau_{c}}$, and the contribution to the excitation population is defined as
    \begin{equation}\label{eq:Pd1}
        P_d^{(1)}(t) \equiv\frac{2(N-1)\sigma^{2}}{N^{2}\Gamma}\left[\frac{1}{q}e^{-N\Gamma t}+\frac{1}{r}-\frac{N\Gamma}{rq}e^{-rt}\right]
    \end{equation}
    Note that $P_d^{(1)}(0)=0$ and 
    \begin{equation}\label{eq:Pd1_infty}
        P_d^{(1)}(t\rightarrow\infty)\rightarrow\frac{2(N-1)\sigma^{2}}{N^{2}\Gamma}\frac{1}{r}\neq0
    \end{equation}
    which yields non-zero population at long times.
\end{enumerate}
At this point, we can put together Eqs.~\eqref{eq:Ps0}, \eqref{eq:Ps1}, \eqref{eq:Pd1} and approximate the excitation population (Eq.~\eqref{eq:Pall}) by
\begin{equation}\label{eq:P_approx}
    P(t)= P_s^{(0)}(t) + P_s^{(1)}(t) + P_d^{(1)}(t). 
\end{equation}
In Fig.~\ref{fig:Kubo}, we compare Eq.~\eqref{eq:P_approx} and the numerical results as obtained by Eqs.~\eqref{eq:electronic_EOM} and \eqref{eq:photon_EOM}. 
We find that, in general, $P_s^{(0)}(t)$ captures the correct superradiant decay at short times, but $P_d^{(1)}(t)$ does not predict the correct subradiant decay at long times. 
That being said, particularly in the parameter region $\gamma<\Gamma$, the time at which $P_s^{(0)}=P_d^{(1)}$ can still provide a good estimation for the critical time $t^*$ at which the population dynamics make a transition from superradiance to subradiance.
\begin{figure}
    \includegraphics{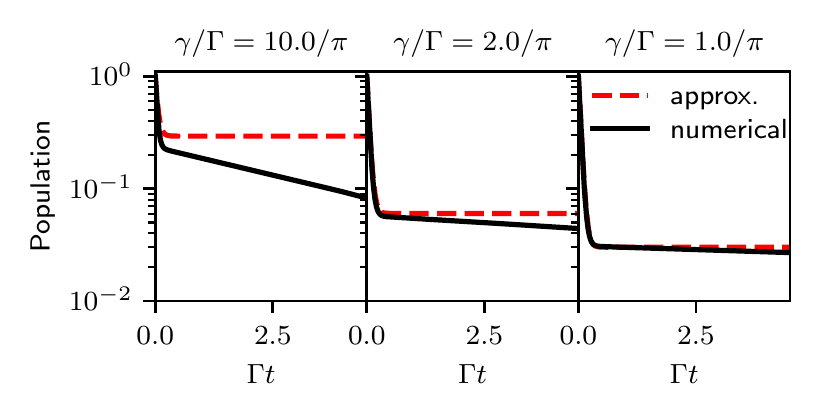}
    \caption{\newparagraph{Molecular excitation population as calculated analytically (using the approximate expression Eq.~\eqref{eq:P_approx}, red dashed line) and as calculated numerically (by brute force, black solid line); the populations  are plotted as a function of time for $\gamma/\Gamma=10/\pi,2/\pi,1/\pi$. 
    %The contributions of $P_{s}^{(0)}$, $P_{s}^{(1)}$, and $P_{d}^{(1)}$ are also plotted as dotted lines separately. 
    We choose $\Gamma=\pi\times10^{-3}$ and $N=20$ as in Fig.~\ref{fig1} and set the initial state to be the superradiant state. The disorder amplitude is fixed $\sigma=0.1$ and the correlation time is $\tau_{c}=1.0,0.2,0.1$, respectively. Note that $P_{s}^{(0)}$ predicts a short-time superradiant decay and $P_{d}^{(1)}$ emerges and dominate for long times. While $P_{d}^{(1)}$ cannot capture the subradiant decay, the approximate population agrees with the numerical result for quite a long time if $\gamma<\Gamma$.}}
    \label{fig:Kubo}
\end{figure}
}

\newparagraph{
\section{Short light pulse excites the superradiant state}
\begin{figure}
    \includegraphics{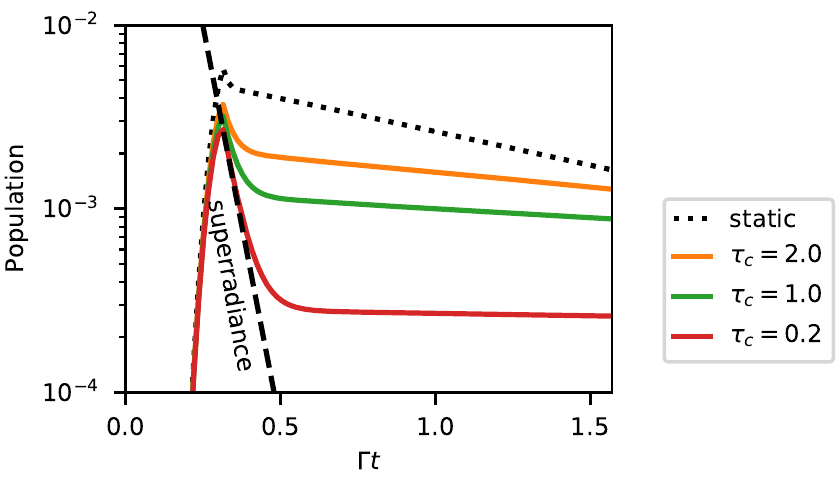}
    \caption{\newparagraph{Molecular excitation population as a function of time under a off-resonant light pulse with a sharp cutoff. After the pulsed excitation, the population dynamics show a superradiant decay at short times followed by a subradiant decay at long times for dynamic disorder. For static disorder, the population dynamics decays at the spontaneous emission rate (black dotted line). Note that dynamic disorder leads to the collective superradiant state.}}\label{fig8}
\end{figure}
We consider a Gaussian light pulse with a sharp cutoff at the peak of the pulse $t_d$:
\begin{equation*}
f(t)=A\sin(\omega_d t)\exp(-(t-t_d)^2/B^2)\left(\frac{1}{2}-\frac{1}{\pi}\arctan(D(t-t_d))\right)
\end{equation*}
Here we choose $A=5\times10^{-3}$, $B=25$, and $D=10^{4}$ for a sharp cutoff. 
The characteristic frequency is off-resonant $\omega_d-\omega_x=0.25$ and $t_d=100$.
In Fig.~\ref{fig8}, we observe that the short pulse excites the collective superradiant state at the peak of the light pulse, then, for dynamic disorder, the excitation population decays at the superradiant rate followed by a subradiant decay. 
In contrast, for static disorder, the superradiant state loses coherence quickly after $t=t_d$ and single-particle emission ensues.
}

\bibliography{references.bib}
\end{document}